\documentclass[fleqn,usenatbib]{mnras}

\usepackage{newtxtext,newtxmath}
\usepackage[T1]{fontenc}
\usepackage{chemformula} % Formula subscripts using \ch{}
\usepackage[T1]{fontenc} % Use modern font encodings
\usepackage{multirow}
\usepackage{bigdelim}
\usepackage{longtable}
\usepackage{graphicx}
\usepackage{times}
\usepackage{amsmath}
\usepackage[utf8]{inputenc}
\usepackage[T1]{fontenc}

\title{FAUST XII. Accretion streamers and jets in the VLA 1623--2417 protocluster}
\author[C. Codella et al.]{C. Codella,$^{1,2}$\thanks{E-mail: claudio.codella@inaf.it}
L. Podio,$^{1}$
M. De Simone,$^{3,1}$
C. Ceccarelli,$^{2}$
S. Ohashi,$^{4}$
C.J. Chandler,$^{5}$
\newauthor
N. Sakai,$^{4}$
J.E. Pineda,$^{6}$
D.M. Segura-Cox,$^{7}$
E. Bianchi,$^{8}$
N. Cuello,$^{2}$
A. L\'{o}pez-Sepulcre,$^{2,9}$
\newauthor
D. Fedele,$^{1}$
P. Caselli,$^{6}$
S. Charnley,$^{10}$
D. Johnstone,$^{11,12}$
Z.E. Zhang,$^{13,4}$
M.J. Maureira,$^{6}$
\newauthor
Y. Zhang,$^{4}$
G. Sabatini,$^{1}$
B. Svoboda,$^{5}$
I. Jim\'enez-Serra,$^{14}$
L. Loinard,$^{15}$
S. Mercimek,$^{1,16}$
\newauthor
N. Murillo,$^{4}$
and S. Yamamoto$^{17,18}$ 
\newauthor
\newauthor
\\
$^{1}$INAF, Osservatorio Astrofisico di Arcetri, Largo E. Fermi 5, I-50125, Firenze, Italy\\
$^{2}$Univ. Grenoble Alpes, CNRS, IPAG, 38000 Grenoble, France\\
$^{3}$European Southern Observatory, Karl-Schwarzschild Str. 2, 85748 Garching bei M\"unchen, Germany\\
$^{4}$RIKEN Cluster for Pioneering Research, 2-1, Hirosawa, Wako-shi, Saitama 351-0198, Japan\\
$^{5}$National Radio Astronomy Observatory, PO Box O, Socorro, NM 87801, USA\\
$^{6}$Max-Planck-Institut f\"ur extraterrestrische Physik (MPE), Gie{\ss}enbachstr. 1, D-85741 Garching, Germany\\
$^{7}$Department of Astronomy, The University of Texas at Austin, 2515 Speedway, Austin, TX 78712, USA\\
$^{8}$Excellence Cluster ORIGINS, Boltzmannstraße 2, 85748, Garching bei M\"unchen, Germany\\
$^{9}$Institut de Radioastronomie Millim\'{e}trique, 38406 Saint-Martin d’H\`{e}res, France\\
$^{10}$Astrochemistry Laboratory, Code 691, NASA Goddard Space Flight Center, 8800 Greenbelt Road, Greenbelt, MD 20771, USA\\
$^{11}$Department of Physics and Astronomy, University of Victoria, 3800 Finnerty Road, Elliot Building Victoria, BC, V8P 5C2, Canada\\
$^{12}$NRC Herzberg Astronomy and Astrophysics 5071 West Saanich Road, Victoria, BC, V9E 2E7, Canada\\
$^{13}$Department of Astronomy, University of Virginia, Charlottesville, VA 22904-4325, USA\\
$^{14}$Centro de Astrobiologia (CSIC-INTA), Ctra. de Torrejon a Ajalvir, km 4, 28850, Torrejon de Ardoz, Spain\\
$^{15}$Instituto de Radioastronomía y Astrofísica , Universidad Nacional Autónoma de México, A.P. 3-72 (Xangari), 8701, Morelia, Mexico\\
$^{16}$Università degli Studi di Firenze, Dipartimento di Fisica e Astronomia, via G. Sansone 1, 50019 Sesto Fiorentino, Italy\\
$^{17}$The Graduate University for Advanced Studies (SOKENDAI), Shonan-village, Hayama, Kanagawa 240-0193, Japan\\
$^{18}$Research Center for the Early Universe, The University of Tokyo, 7-3-1 Hongo, 
Bunkyo-ku, Tokyo 113-0033, Japan \\
}

\date{Accepted XXX. Received YYY; in original form ZZZ}
\date{Accepted XXX. Received YYY; in original form ZZZ}

\pubyear{2022}

\begin{document}
\label{firstpage}
\pagerange{\pageref{firstpage}--\pageref{lastpage}}
\maketitle

\begin{abstract}
%The advent of the ALMA interferometer, with its high sensitivity, high angular resolution, and large spectral coverage, boosted the investigation of Sun-like protostellar systems. More precisely, 
The ALMA interferometer has played a key role in revealing a new component of the Sun-like star forming process: the  molecular streamers, i.e. structures up to thousands of au long funneling material non-axisymmetrically to disks. In the context of the FAUST ALMA LP, 
the archetypical VLA1623-2417 protostellar cluster has been imaged at 1.3 mm in the SO(5$_6$--4$_5$), SO(6$_6$--5$_5$), and SiO(5--4) line emission at the spatial resolution of 50 au. We detect extended SO emission, peaking towards the A and B protostars.
Emission blue-shifted down to 6.6 km s$^{-1}$ reveals for the first time a long ($\sim$ 2000 au) accelerating  streamer plausibly feeding the VLA1623 B protostar. 
%Furthemore, we confirm the occurrence of a smaller (250 au) northern streamer, impacting the A circumbinary disk.
Using SO, we derive for the first time an estimate of the excitation temperature of an accreting streamer:
%, in this case the southern streamer: 
33$\pm$9 K. The SO column density is $\sim$ 10$^{14}$ cm$^{-2}$, and the SO/H$_2$ abundance ratio is $\sim$ 10$^{-8}$.
The total mass of the streamer is 3 $\times$ 10$^{-3}$ $M_{\rm \sun}$, while its accretion rate is 3--5 $\times$ 10$^{-7}$ $M_{\rm \sun}$ yr$^{-1}$.
This is close to the mass accretion rate of VLA1623 B, in the 0.6--3 $\times$ 10$^{-7}$ $M_{\rm \sun}$ yr$^{-1}$ range, showing the importance of the streamer in contributing to the mass of protostellar disks.
The highest blue- and red-shifted SO velocities behave as the SiO(5--4) emission, the latter species detected for the first time in VLA1623-2417:
the emission is compact (100-200 au), and associated 
only with the B protostar. The SO excitation temperature 
is $\sim$ 100 K, supporting the occurrence of 
%92$\pm$18 K (red-shifted velocities) and 102$\pm$19 K (blue-shifted), 
shocks associated with the jet, traced by SiO. 
\end{abstract}

%%%%%%%%%%%%%%%%%%%%%%%%%%%%%%%%%%%%%%%%%%%%%%%%%%
%%%%%%%%%%%%%%%%% BODY OF PAPER %%%%%%%%%%%%%%%%%%

\section{Introduction} \label{sec:intro}

%\subsection{The formation of a Sun-like star}

The low-mass star forming process takes a dense core of gas and dust inside a molecular cloud and leaves a 
Sun-like star possibly surrounded by its planetary system
\citep{shu1977}. Historically, the youngest two classes of Sun-like protostars have beem classified in Class 0 and Class I objects
\citep{Andre1993,Andre2000}, being 10$^{4}$ yr and 10$^{5}$ yr old, respectivey. The standard picture shows a collapse of the slowly infalling envelope (spatial scale of $\sim$ 1000 au) accreting the protostellar mass through a rotating equatorial accretion disk ($\sim$ 50 au). At the same time, angular momentum is removed via fast ($\sim$ 100 km s$^{-1}$) jets ejected from both the protostellar poles \citep[e.g.][]{Terebey1984,Frank2014,Lee2020}, pushing in turn slower outflows. 
All these physical components, characterised by different velocities, have been imaged using the proper combination between spatial resolution and molecular tracer \citep[e.g.][and references therein]{Codella2019,Ceccarelli2023}. As an example, envelopes 
and outflows can be well traced by CO and its rarer isotopologues, while the classical jet tracers 
are the SiO isotopologues. The so-called interstellar Complex Organic Molecules \citep[iCOMS, i.e. organic species with at least 6 atoms such as CH$_3$OH,][and references therein]{Herbst2009,Ceccarelli2023} are able to trace the inner 100 au around the protostars where the temperature is high enough ($\geq$ 100 K) to release species from dust mantles to the gas phase. Finally, the protostellar disk has been traced by the chemical enrichment (iCOMs, S-bearing species, such as SO) due to mild shocks occurring near the centrifugal barrier, where the infalling envelope has to lose energy to proceed on its journey to the protostar through the accretion disk \citep[][and references therein]{Sakai2014a,Sakai2014b,Oya2016,Lee2019,Ceccarelli2023}.

As matter of fact, the classic protostellar collapse picture predicted axisymmetry of the protostellar structures with respect to the disk equatorial plane, and/or the jet
axis \citep[e.g.]{Frank2014}, which was generally 
supported by observations until recently.  Nonetheless, a new component has been detected thanks to high sensitivity interferometric images: the molecular streamers, i.e. elongated structures revealed in the protostellar environment, which could significantly contribute to the mass accretion of the newly born stars
\citep[see the recent review by][and references therein]{Pineda2023}. Using IRAM-NOEMA, \citet{Pineda2020} discovered the presence of a large scale (10000 au) accretion streamer in HC$_3$N line emission towards 
the Class 0 object Per-emb-2. Successively, other streamers
(as long as 6000 au) have been imaged around well known Class 0 protostars with ALMA: Lupus3-MMS \citep[CO isotopologues,][]{Thieme2022}, and IRAS16293--2422 A \citep[HNC, HC$_3$N, HC$_5$N,][]{Murillo2022}. Thanks to ALMA, accretion streamers have been detected also towards more evolved Class I objects,
starting from the archetypical HL Tau disk, where \citet{Yen2019} imaged in HCO$^{+}$(3--2) a 200 au structure rotating and infalling to the disk. 
In addition, (i) \citet{Garufi2021} imaged a small ($\sim$ 150 au) CS(5--4) streamer towards DG Tau, while (ii) \citet{Garufi2022} and \citet{Bianchi2023} showed evidence for shocks due to the encounter between disks and infalling streamers in DG Tau, HL Tau, and SVS13-A using SO, SO$_2$, and HDO emission. 
Finally, using IRAM-NOEMA, \citet{Valdivia2022} revealed a long ($\geq$ 2000 au) streamer in HCO$^{+}$ and C$^{18}$O in the Class I object Per-emb-50,
while \citet{Hsieh2023} imaged a DCN and C$^{18}$O streamer 700 au long accreting onto the SVS13-A binary.

In summary, there is evidence that molecular streamers are funneling fresh material in an asymmetric way towards protostellar disks at the earliest protostellar phases. This is even more important taking into account that one of the main ALMA breaktrough results is that planet formation may start already around protostars less than 1 Myr old \citep[e.g.][]{Sheehan2017,Fedele2018,Segura2020}. The investigation of molecular streamers has just started: the more efficient molecular tracers have not been identified yet, as well as the typical lenghts of the elongated structures. More observations are clearly needed to draw a more detailed picture \citep{Pineda2023}.
In this paper, in the context of the ALMA Large program
FAUST \footnote{http://faust-alma.riken.jp; \citet{Codella2021}} (Fifty AU STudy of the chemistry in the disk/envelope system of Solar-like protostars), we present a survey in SO and SiO of the VLA 1623--2417 protostellar cluster in order to reveal material associated with accretions streamers as well as protostellar jets.

\section{The VLA1623--2417 protostellar system} \label{sec:target}

The VLA 1623--2417 (hereafter VLA 1623) region, located in Ophiucus A  \citep[d = 131$\pm$1 pc,][]{Gagne2018} is one of the most studied protostellar systems in the Southern hemisphere. 
VLA 1623 has several components imaged at different spectral wavelengths \citep[e.g.][and references therein]{Andre1990,Andre1993,Leous1991,Looney2000,Ward2011,Murillo2013disk,Murillo2018L,Murillo2018,Harris2018,Hsieh2020,Ohashi2022,Codella2022,Mercimek2023}: (i) a binary system
made up of two Class 0 objects, A1 and A2, separated by less than 30\,au, and surrounded by a circumbinary disk;
(ii) another Class 0, labelled B, lies outside of the A1+A2 circumbinary disk, at a projected angular separation of $\simeq$ 1$\arcsec$ ($\sim$130\,au);
in addition, a more evolved Class I object, labelled W, is located at $\sim$ 1200\,au west of the VLA1623 A1+A2+B system.

\begin{figure*}
\begin{center}
\includegraphics[scale=0.5]{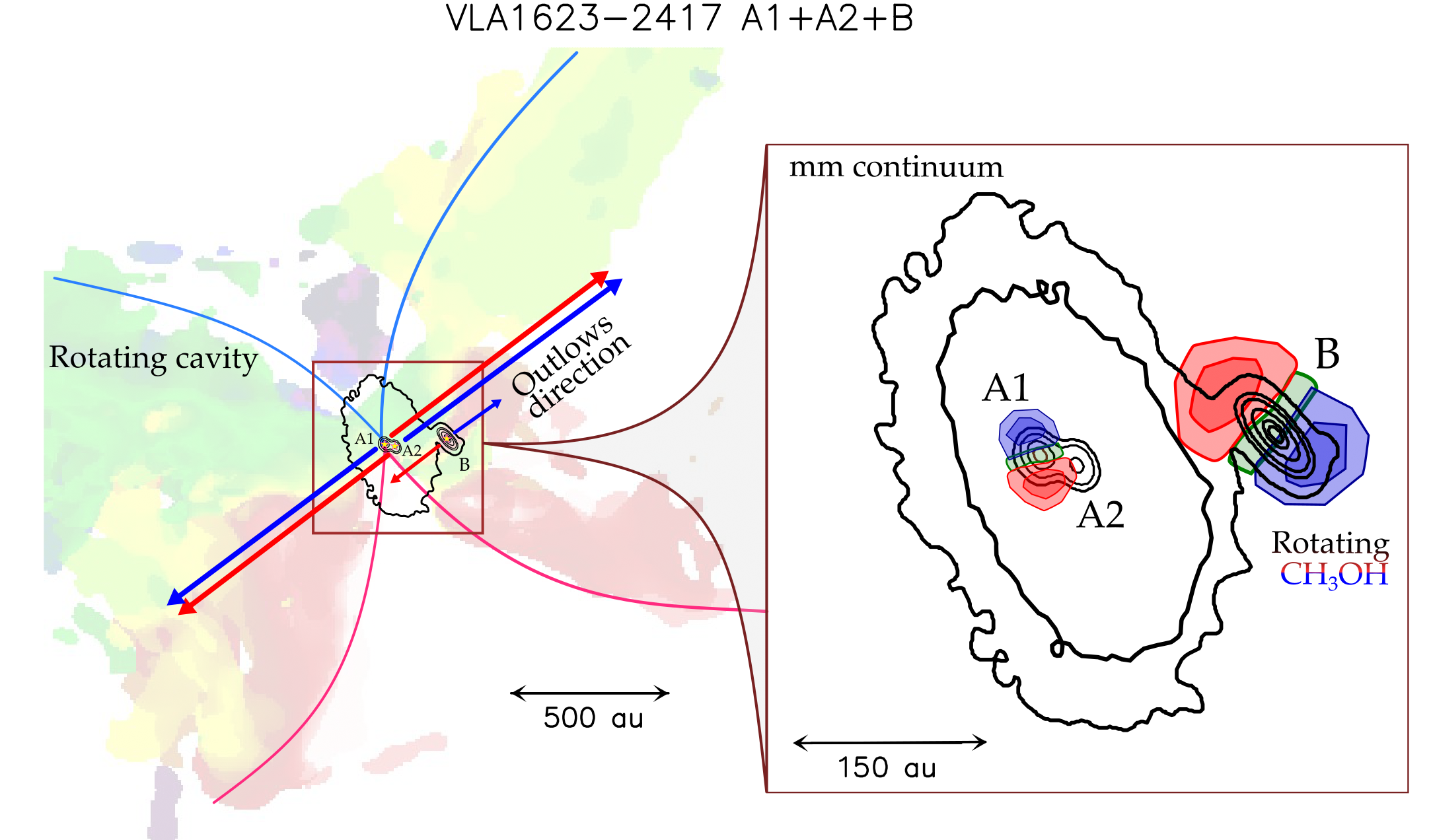}
\caption{Sketch (not to scale) of the VLA1623--2417 system \citep[see also Fig. 19 by][]{Hsieh2020}.
The figure shows: 
(i) the high-spatial resolution mm-continuum map \citep{Harris2018}
revealing the A1+A2 binary system, its circumbinary disk, and the protostar B, (ii) the extended rotating cavity \citep[CS,][]{Ohashi2022}, (iii) the directions of the multiple outflows \citep[CO,][]{Santangelo2015,Hsieh2020,Hara2021}, and (iv) the rotating disks of A1 and B imaged in CH$_3$OH \citep{Codella2022}.}
\label{fig:sketch}
\end{center}
\end{figure*}

Given its complexity, the VLA 1623 star forming region is a perfect laboratory to study the interaction of the star forming process with the surrounding medium. Figure \ref{fig:sketch} provides a sketch (not to scale) summarising some processes imaged in VLA1623 and discussed here \citep[see also Fig. 19 by][]{Hsieh2020}. 
From the kinematic point of view, three main processes have been detected: (1) outflowing motion, (2) gravitationally supported disks, and (3) infalling molecular streamers. 
These processes are described further below.

\begin{enumerate}
\item[(1)] Extended ($>$ 1000 au) outflows along a NW-SE direction have been observed in a number of species (e.g. CO isotoplogues) driven by the A+B multiple system \citep[e.g.][and references therein]{Andre1990,Caratti2006,Hsieh2020,Hara2021}. \citet{Santangelo2015} imaged a fast CO jet from VLA1623 B, while the number of flows driven by A1+A2 is controversial. On the one hand, \citet{Hsieh2020} and \citet{Hara2021} reported two cavities opened by two outflows along the same projected NW-SE direction driven by A1 and A2. As part of ALMA-FAUST, \citet{Ohashi2022} sampled (with a beam of 50\,au) a unique, rotating, and low-velocity NW-SE cavity opened by A1; 

\item[(2)] C$^{18}$O, CS, and CH$_3$OH emission around both VLA1623-2417 A1 and B shows velocity gradients (on 20-30 au scale) along the NE-SW direction \citep{Murillo2013disk,Ohashi2022,Codella2022}, i.e. along the main axis of each protostellar disk observed in continuum \citep{Harris2018};

\item[(3)] Recently, the occurrence of molecular
streamers have been reported by \citet{Hsieh2020} imaging SO(8$_{\rm 8}$-7$_{\rm 7}$)
at a spatial scale $\sim$ 100 au. The authors
support the occurrence of 
two blue-shifted northern flows accreting onto both the circumbinary disk around the A binary and the B protostellar disk, plus a red-shifted southern flow feeding B from the SW direction.
The largest recoverable scale of the SO maps by \citet{Hsieh2020} is 3$\farcs$5, calling for further observations imaging more lines 
and larger spatial scales to confirm the occurrence of extended accretion streamers.
\end{enumerate}

\section{Observations} \label{sec:obs}

The VLA1623 multiple system was observed between 2018 December, and 2020 March with ALMA Band 6 
(Setup 1: 214.0--219.0\,GHz and 229.0--234.0\,GHz, Setup 2: 242.5--247.5\,GHz and 257.2--262.5\,GHz) in the context of the FAUST Large Program (2018.1.01:205.L, PI: S. Yamamoto), using the 12-m array (C43-1, C43-4, with 48 and 49 antennas, respectively) as well as the ACA (Atacama Compact Array) 7-m array (12 antennas). The baselines were between 9\,m ($B_{\rm min}$) and 969\,m ($B_{\rm max}$), for a maximum recoverable scale ($\theta_{\rm MRS}$\,$\sim$\,$0.6\,\lambda\,B_{\rm min}^{-1}$) of $\sim\,$29$\arcsec$. The observations were centered at $\alpha_{\rm J2000}$ = 16$^{\rm h}$\,26$^{\rm m}$\,26$\fs$392, $\delta_{\rm J2000}$ = --24$\degr$\,24$\arcmin$\,30$\farcs$178. The lines here analysed are SO(5$_{\rm 6}$--4$_{\rm 5}$) (219.9 GHz), SO(6$_{\rm 6}$--5$_{\rm 5}$) (258.3 GHz), and
SiO(5--4) (217.1 GHz): the spectroscopic parameters are reported in Table 1. The SO and SiO lines were observed using spectral windows with a bandwidth/frequency resolution of
59 MHz/122 kHz ($\sim$80\,km s$^{-1}$/0.14--0.17 km s$^{-1}$).
The FWHM Field of View (FoV) of the ALMA images are:
26$\arcsec$ for SO(5$_6$--4$_5$) and SiO(5--4), and 22$\arcsec$ for SO(5$_6$--4$_5$). A wide bandwidth (1.875 GHz) spectral window was also included to support measurement of the continuum emission.
Data were calibrated using the quasars J1427-4206, J1517-2422, J1625-2527, J1924-2914, and J1626-2951, reaching an absolute flux calibration uncertainty of $\sim$10\%. The data were self-calibrated using line-free continuum channels. The primary beam correction has also been applied. We used the calibration pipeline\footnote{https://github.com/autocorr/faust$\_$line$\_$imaging; Chandler et al. (in preparation)} within \textsc{CASA 5.6.1-8} \citep{CASA2022}, including an additional calibration routine to correct for $T_{\rm sys}$ issues and spectral data normalization\footnote{https://help.almascience.org/kb/articles/what-errors-could-originate-from-the-correlator-spectral-normalization-and-tsys-calibration; Moellenbrock et al. (in preparation)}. As a consequence, the dynamical range of the continuum data improved up to one order of magnitude. Once the three array configurations were merged, the resulting continuum-subtracted line-cubes were cleaned with a Briggs parameter of 0.5. The data analysis was performed using the \textsc{IRAM-GILDAS}\footnote{http://www.iram.fr/IRAMFR/GILDAS} package. The continuum 
has been imaged using uniform weighting, thus obtaining a beam of 0$\farcs$37 $\times$ 0$\farcs$34 ($-65^\circ$), and 0$\farcs$43 $\times$ 0$\farcs$32 ($-65^\circ$), for Setup 1 and Setup 2, respectively.
On the other hand, the r.m.s. noise is 0.22 mJy beam$^{-1}$ (Setup 1), and 0.15 mJy beam$^{-1}$ (Setup 2).
The synthesized beams of the line datasets are 0$\farcs$54$\times$0$\farcs$45 (PA=--75$^{\circ}$), for Setup 1, and 0$\farcs$48$\times$0$\farcs$45 (PA=+86$^{\circ}$), for Setup 2. The typical r.m.s. noise (per channel) is $\sim$3 mJy beam$^{-1}$. To decrease the noise, the SiO(5--4) datacube has been  spectrally smoothed to 1 km s$^{-1}$, for an r.m.s. of 1 mJy beam$^{-1}$.

\begin{figure*}
\begin{center}
\includegraphics[scale=0.6]{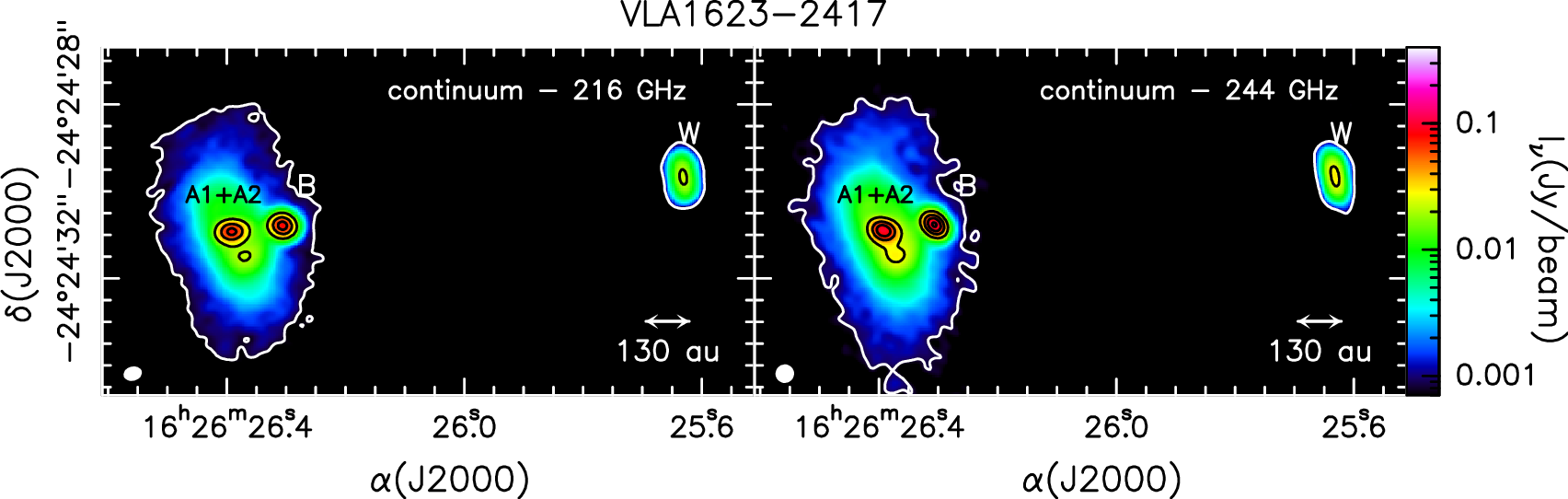}
\caption{Dust continuum emission at 216 GHz and 244 GHz (colour scale and contours) from the VLA1623-2417 multiple system. First contours, in white, are 3$\sigma$ (0.8 mJy beam$^{-1}$). Steps are 100$\sigma$. The synthesised beam (bottom-left corners) are 0$\farcs$43 $\times$ 0$\farcs$32 (PA = --65$\degr$), and 0$\farcs$38 $\times$ 0$\farcs$35 (PA = +66$\degr$), for the 216 GHz and 244 GHz maps, respectively. The A1 and A2 protostars are not disentangled at the present angular resolutions. The B and W protostars are also labelled.} \label{fig:continuum}
\end{center}
\end{figure*}

\begin{table}
\caption{Spectral Properties of the SO and SiO lines observed towards VLA1623.} 
\label{Tab:lines}
\begin{tabular}{lcccc}
\hline
Transition & $\nu^{\rm a}$ & E$_{\rm u}^{\rm a}$ & Log$_{\rm 10}$(A$_{\rm ul}$/s$^{-1})^{\rm a}$ & $S\mu^{\rm 2a}$ \\ 
 & (MHz) & (K) &  & (D$^2$)  \\
\hline
SO(5$_{\rm 6}$--4$_{\rm 5}$) & 219949.442 & 35 & --3.9 & 14.0 \\
SO(6$_{\rm 6}$--5$_{\rm 5}$) & 258255.826 & 57 & --3.7 & 13.7 \\
SiO(5--4) & 217104.980 & 31 & --3.3 & 48.0 \\
\hline
\end{tabular}
\\
$^{\rm a}$ Spectroscopic parameters are from \citet{Klaus1996}, and \citet{Bogey1997} (SO), and \citet{Manson1977}, for SiO, retrieved from the CDMS database \citep{Muller2005}. \\
\end{table}

\section{Results} \label{sec:results}

\subsection{Continuum emission}

Figure \ref{fig:continuum} shows the VLA 1623 region as observed in dust continuum emission at 216 GHz (1.4 mm) and 244 GHz (1.2 mm). 
A 1.2 mm image has been already reported in the context of the FAUST campaign by \citet{Codella2022}, but only using the C43-4 configuration of the 12m array.
%The spatial resolution was in that case 0$\farcs$39 $\times$ 0$\farcs$6 and the largest recoverable scale $\sim$ 45$\arcsec$ lower than that of the present dataset including ACA ($\sim$ 70$\arcsec$).
The 1.4 mm image has been presented by \citet{Mercimek2023}
in the context of the analysis of source W.

The FAUST continuum images show the envelope containing 
the A1 and A2 binary system (not disentangled by the present spatial resolution) and the B protostar. The emission from
the A1+A2 circumbinary disk is also revealed. At about 1300 au west of the A+B system, the W protostar is also detected.
The J2000 coordinates of the A, B, and W protostars, as traced by the 2D fitting of 
both the 1.2 mm and 1.4 mm images 
are A: 16$^{\rm h}$\,26$^{\rm m}$\,26$\fs$392, --24$^\circ$\,24$'$\,30$\farcs$90;
B: 16$^{\rm h}$\,26$^{\rm m}$\,26$\fs$307, --24$^\circ$\,24$'$\,30$\farcs$76;
W: 16$^{\rm h}$\,26$^{\rm m}$\,25$\fs$632, --24$^\circ$\,24$'$\,29$\farcs$64.
In summary, the FAUST picture is well in agreement with 
the ALMA image at 0.9 mm obtained by \citet[][see also the references therein]{Harris2018} with a resolution of 0$\farcs$2. A detailed analyis of continuum emission is beyond the scope of this paper. Continuum images will contribute to the analysis of the origin of the SO and SiO (Sect. 4) gas observed in the A+B system.

\subsection{Overall SO spatial distribution and spectra}

\begin{figure*}
\begin{center}
\includegraphics[angle=0,scale=0.48]{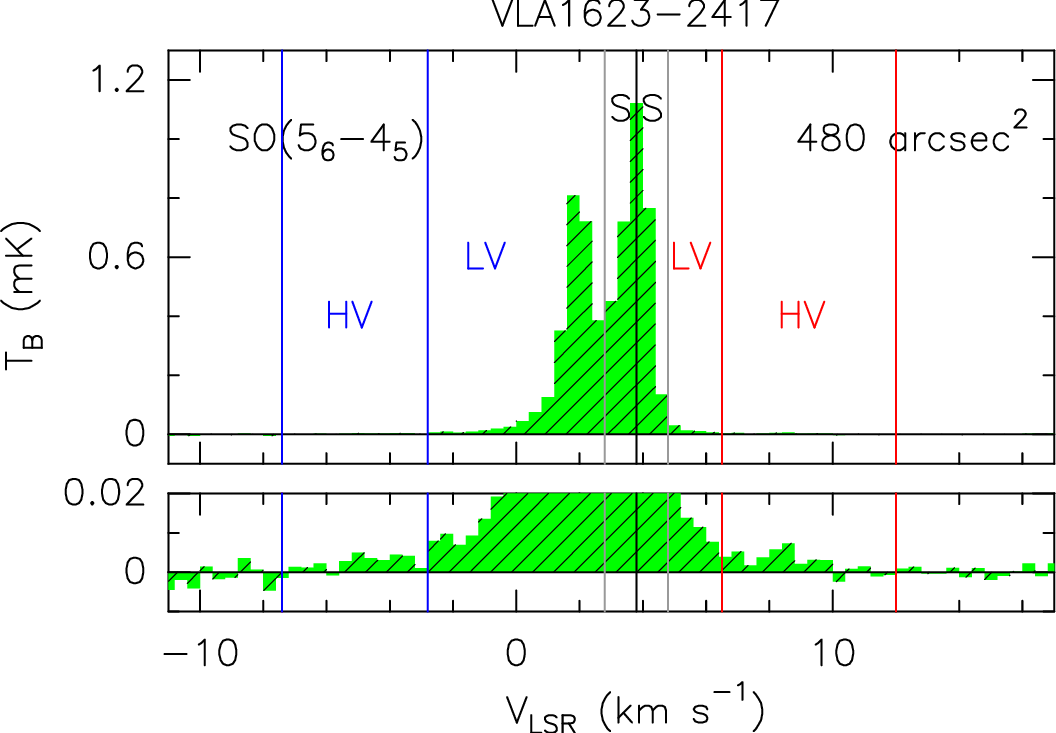}
\includegraphics[angle=0,scale=0.48]{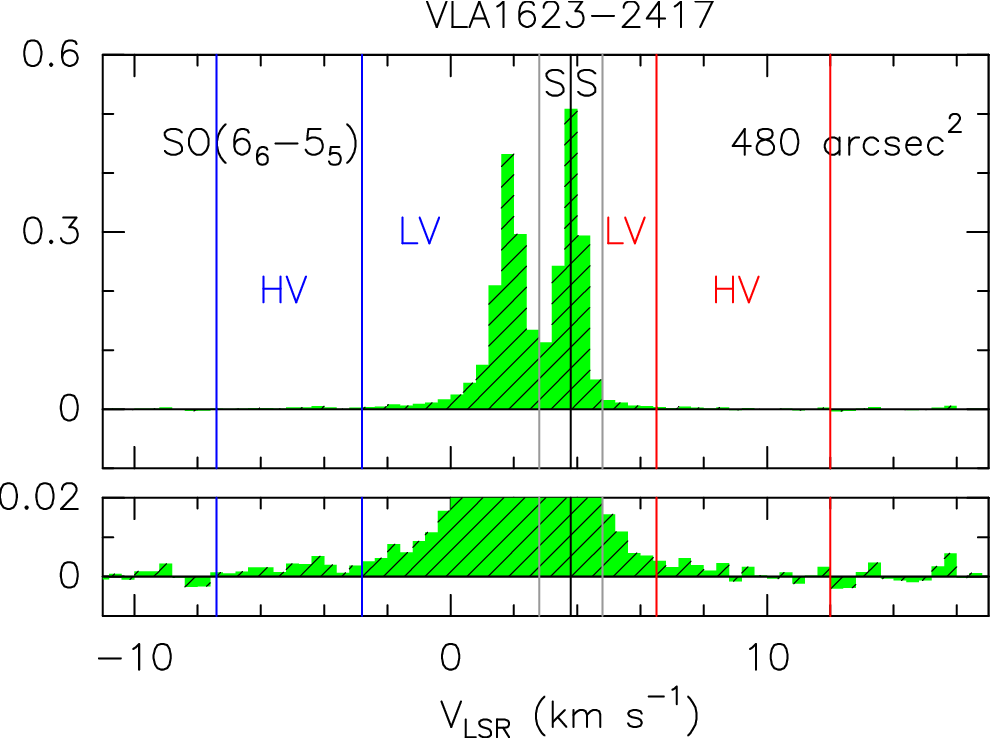}
 \caption{{\it Upper panels:} SO(5$_6$--4$_5$) (Left), and SO(6$_6$--5$_5$) (Right) spectra (in brightness temperature, $T_{\rm B}$, scale) derived integrating the emission over 480 arcsec$^2$, i.e. a region 22$\arcsec$ wide centred around the A1+A2+B protostars (see Fig. \ref{fig:mom0}). In both panels, the brightness temperature r.m.s. is $\sim$ 1 mK. The black vertical line is for the systemic velocity of the triple system of $V_{\rm sys}$ = +3.8 km s$^{-1}$ \citep{Ohashi2022}. The grey vertical lines show the velocity range $\pm$ 1 km s$^{-1}$ with respect to $V_{\rm sys}$ (labelled S). The blue and red vertical lines 
 delimitate the blue- and red-shifted velocity ranges 
 tracing different SO structures, as described in Sect. 3. More precisely, the
 velocity range with a shift between 1.0 km s$^{-1}$ 
 and 6.6 km s$^{-1}$ (blue) or 2.5 km s$^{-1}$ (red) 
 is labelled as LV. The label HV is for the highest and weakest SO emission (see the results on kinematics of Sect. 4). {\it Bottom panels:} Zoom-in of the same SO spectra of the Upper panels shown to highlight the weak high-velocity emission.} \label{fig:spectraSO}
\end{center}
\end{figure*}

Both SO(5$_6$--4$_5$) and SO(6$_6$--5$_5$) emission lines have been detected and imaged. Figure \ref{fig:spectraSO} shows the SO(5$_6$--4$_5$) and SO(6$_6$--5$_5$) line profiles derived integrating the emission over a region as large as the Field of View of the SO map at 258 GHz (22$\arcsec$).
In Fig. \ref{fig:spectraSO} (Bottom panels)
the zoom-in shows the weakest SO emission, offset in velocity up to $\sim$ 10 km s$^{-1}$ with respect to the systemic velocity of the A+B system, i.e. $V_{\rm sys}$ = +3.8 km s$^{-1}$ \citep{Ohashi2022}. More precisely, the velocity range goes from --7.6 to +12.0 km s$^{-1}$.
Emission due to SO has been recently reported by \citet{Hsieh2020}, who detected the SO(8$_8$--7$_7$) line at 344 GHz with ALMA, in a narrower velocity range, from $\sim$ --2 km s$^{-1}$ to $\sim$ +6 km s$^{-1}$.
Figure \ref{fig:mom0} reveals the spatial distribution
of the SO(5$_6$--4$_5$) and SO(6$_6$--5$_5$) emission as integrated over the whole emitting velocity range (moment 0 maps). The present SO maps improve the spatial resolution of the image collected by \citet{Hsieh2020}, obtained with a synthetised beam of 1$\farcs$11 $\times$ 0$\farcs$76. 
Figure \ref{fig:spectraSOSiO} reports the SO(5$_6$--4$_5$) and SO(6$_6$--5$_5$) spectra extracted at the positions of the A, and B peaks.
Emission is also detected towards the object W, located at the edge of the FoV of the SO(6$_6$--5$_5$) image, but its analysis is out of the scope of the present paper. Those maps show that SO has compact emission peaking on
A and B, but also shows extended and elongated structures not associated with the VLA1623 outflows. Both components are discussed below. 

\subsection{SO emission close to the A and B protostars}

The close association of the SO peaks with the protostellar positions 
suggest a possible contribution from hot-corinos, where the temperature is high
enough ($\geq$ 100 K) to allow evaporation into the gas phase of the icy mantles. Recent observations \citep{Codella2022} of VLA1623-2417 imaged methanol emission rotating, on small-spatial scales, around the protostars A1 and B (see Fig. \ref{fig:sketch}). 
%More precisely, the LVG analysis of the emission around the B protostar indicated kinetic temperatures larger than 180 K in a size of 45 au. 

The linewidth of the SO spectra extracted at the A continuum peak (see Figure \ref{fig:spectraSOSiO}) is 1.8 km s$^{-1}$, narrower than the 4 km s$^{-1}$ methanol profile observed by \citet{Codella2022}. 
However, the entire SO lines are broader, and they look affected by absorption at velocities close to $V_{\rm sys}$, more specifically at slightly red-shifted velocities, as found observing CS(5--4) by \citet{Ohashi2022}. As a consequence, the contribution of the SO emission by the hot-corino in A cannot be assessed using the observed lines.

The line profiles extracted at the B continuum peak (Fig. \ref{fig:spectraSOSiO}) protostar are more complex: the lines are very broad ($\sim$ 8 km s$^{-1}$) with, in addition, extended wings suggesting the occurrence of a high-velocity jet. An absorption dip is observed at velocities close to the systemic one in the SO(5$_6$--4$_5$) line, whereas a weak absorption is present in the SO(6$_6$--5$_5$) profile. A remarkable absorption along the line of sight of B, down to negative brightness temperatures, has been observed by \citet{Ohashi2022} using CS, CCH, and H$^{13}$CO$^{+}$ low-excitation ($E_{\rm u}$ = 25--35 K) lines. Those profiles suggest absorption against an optically thick continuum in the background, associated with the protostar.
The present SO profiles are also consistent
with material placed between the material surrounding the protostars and the observer.
As shown in Fig. \ref{fig:spectraSOSiO}, the absorption is more prominent for the SO(5$_6$--4$_5$) line ($E_{\rm u}$ = 35 K) with respect to the SO(6$_6$--5$_5$) one ($E_{\rm u}$ = 57 K), suggesting low-excitation absorbing material or an optically thick continuum. 

%Finally, Figure \ref{fig:mom0} (Bottom panels) shows the SO
%profiles extracted at the position of the protostar W, characterised by two peaks, red- and blue-shifted by about 1.5 km s$^{-1}$ with respect to the systemic velocity,
%which is +1.6 km s$^{-1}$ for the protostar W \citep{Ohashi2022}. 
%The spectra are in agreement with the recent analysis of 
%\citet{Mercimek2023},
%who mapped the W object using C$^{18}$O emission, finding gas rotation 
%at velocities larger than $\pm$ 1.8 km s$^{-1}$ with respect to the systemic.
%\citet{Mercimek2023} interpreted this velocity pattern as a disk or the inner ($\leq$ 50 au) portion of the protostellar envelope.

\subsection{Extended SO emission}

\begin{figure*}
\begin{center}
\includegraphics[angle=0,scale=0.48]{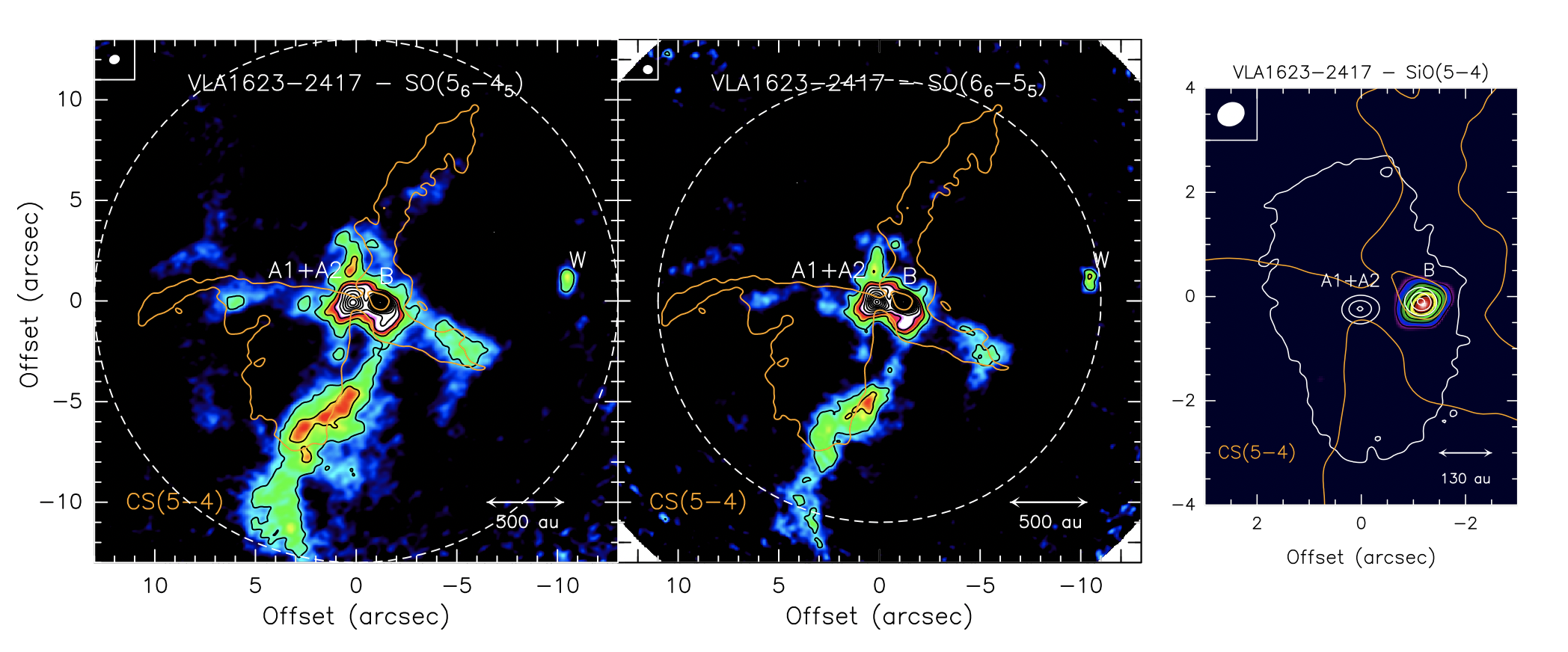}
 \caption{The VLA1623--2417 system as traced by the integrated intensity map (moment 0, color scale and contours) of SO(5$_6$--4$_5$) (Left panel),
 SO(6$_6$--5$_5$) (Middle), and SiO(5--4) (Right). 
 %The position of the A1+A2, B, and W protostars are labelled.
The SO emission is integrated from --7.6 to +12.0 km s$^{-1}$, while that
of SiO map from +0.6 to +5.2 km s$^{-1}$.
First contours of both the SO maps start from 3$\sigma$ (27 mJy km s$^{-1}$ beam$^{-1}$) with intervals of 9$\sigma$. 
First contour of the SiO image start from 5$\sigma$ (10 mJy km s$^{-1} $ beam$^{-1}$) with intervals of 3$\sigma$. 
The synthesised beam (top-left corners) are 0$\farcs$54 $\times$ 0$\farcs$45 (PA = --74$\degr$), for SO(5$_6$--4$_5$) and SiO(5--4), and 0$\farcs$47 $\times$ 0$\farcs$45 (PA = +86$\degr$), for SO(6$_6$--5$_5$).
The dashed circles delimitate the FWHM Field of View of the ALMA image:
26$\arcsec$ for SO(5$_6$--4$_5$) and SiO(5--4), and 22$\arcsec$
for SO(5$_6$--4$_5$).
In the Right panel, white contours representing selected intensities of the continuum map at 216 GHz (see Fig. \ref{fig:continuum}) are drawn to
show the position of the A1+A2, and B protostars.
The orange thick contour is the CS(5--4) emission (25$\sigma$) which traces the outflow cavity walls associated with VLA1623A1+A2
\citep[from][]{Ohashi2022}.} \label{fig:mom0}
\end{center}
\end{figure*}

\begin{figure*}
\begin{center}
\includegraphics[angle=0,scale=0.48]{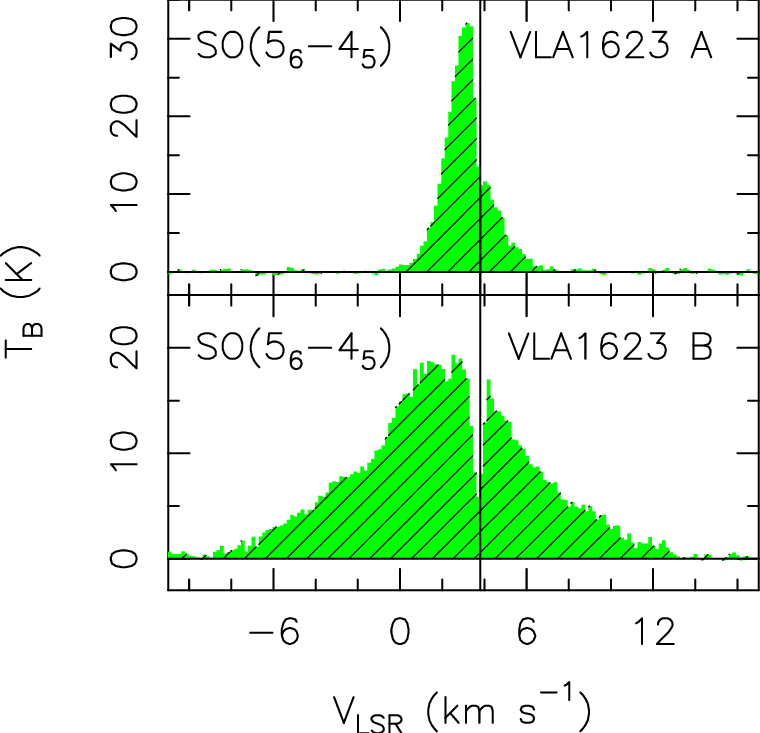}
\includegraphics[angle=0,scale=0.48]{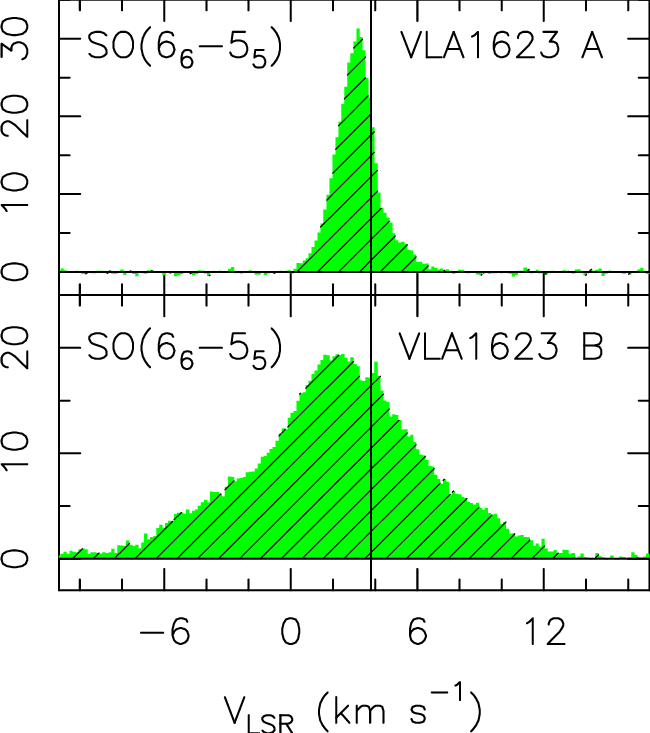}
\includegraphics[angle=0,scale=0.48]{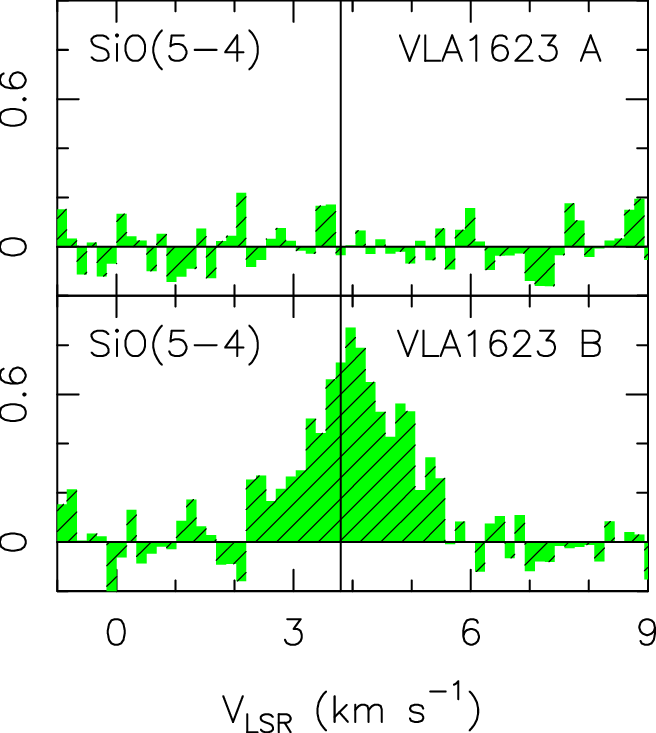}
 \caption{SO(5$_6$--4$_5$) (Left-panels), and SO(6$_6$--5$_5$) (Middle),
 and SiO(5--4) (Right) spectra (in brightness temperature, $T_{\rm B}$, scale) derived at the two peaks of the continuum
 maps: A (Upper), and B (Bottom), see Fig. \ref{fig:continuum}. The black vertical lines are for the systemic velocity, i.e. +3.8 km s$^{-1}$  \citep{Ohashi2022}.} \label{fig:spectraSOSiO}
\end{center}
\end{figure*}

The elongated SO structures can be compared with the spatial distribution of the CS cavities (orange contours in Fig. \ref{fig:mom0}), opened by the outflow located along the NW-SE direction and driven by the VLA1623A1 object \citep{Ohashi2022}.
Figure \ref{fig:mom0} shows that two elongated structures lie outside the static CS cavities: (i) a very long ($\sim$ 1500 au) one in the region south of the multiple protostellar system, and (ii) one located north of A1+A2, $\sim$ 250 au long.
The present large scale picture shows some differences with respect to that drawn by \citet{Hsieh2020} using the SO(8$_8$--7$_7$) line: on the one hand we confirm the occurrence of the elongated structure north of A1+A2; on the other hand, the SW emission looks associated with the molecular cavity.
%while no signature of an accretion flow feeding B from the NW direction \citep[called Accretion Flow III by][]{Hsieh2020} is here revealed. 

\subsubsection{Southern region: the VLA1623 B accretion streamer}

\begin{figure}
\begin{center}
\includegraphics[scale=0.69]{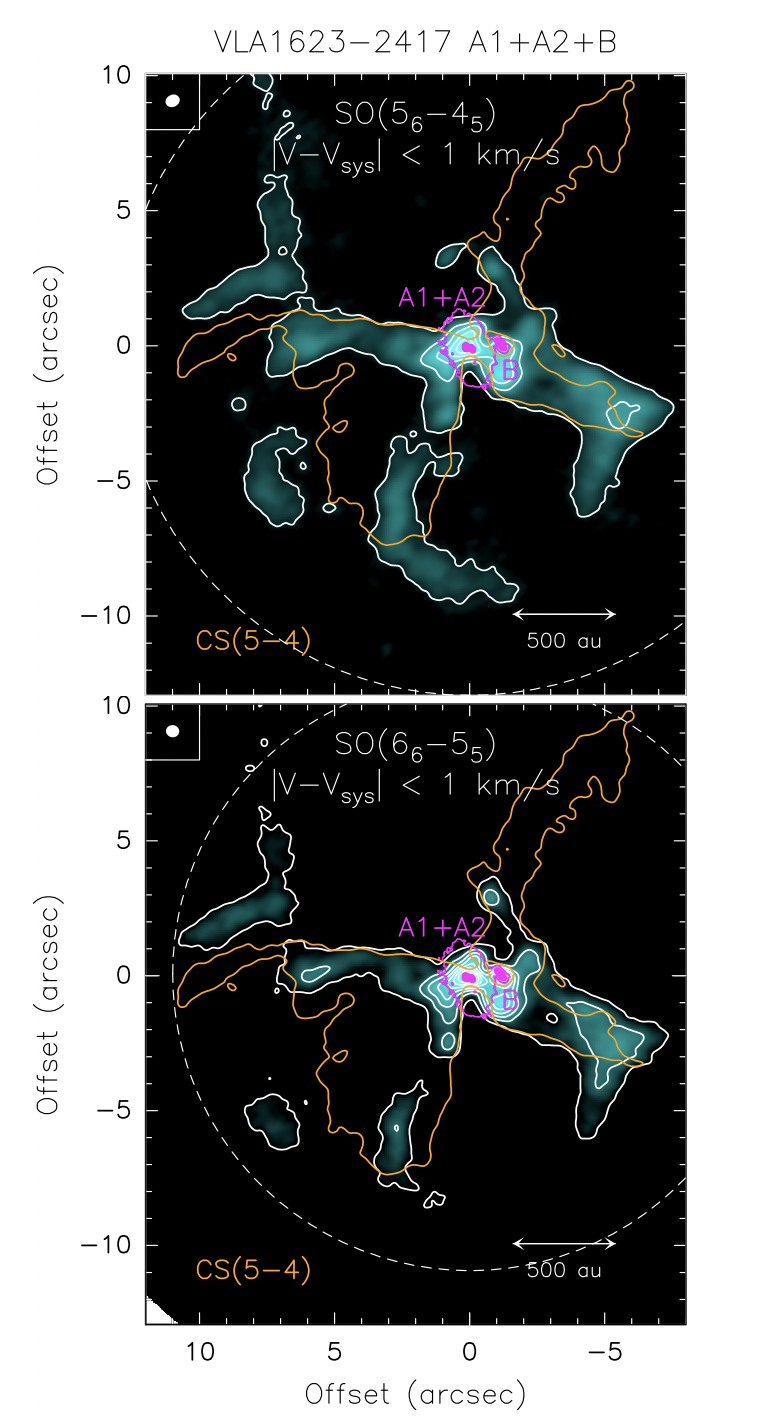}
\caption{The VLA1623--2417 A1+A2+B system as traced by SO(5$_6$--4$_5$) (Upper panel),
 and SO(6$_6$--5$_5$) (Bottom) emission integrated over $\pm$ 1 km s$^{-1}$ (labelled S in Fig. \ref{fig:spectraSO}) with respect to the systemic velocity of the triple system of +3.8 km s$^{-1}$ \citet{Ohashi2022}. The position of the A1+A2, B, and W protostars are labelled. The dashed circles delimitate the FWHM Field of View of the ALMA image: 26$\arcsec$ for SO(5$_6$--4$_5$), and 22$\arcsec$ for SO(5$_6$--4$_5$). The dashed circles delimitate the FWHM Field of View of the ALMA image. The synthesised beam (top-left corners) are 0$\farcs$54 $\times$ 0$\farcs$45 (PA = --74$\degr$), for SO(5$_6$--4$_5$), and 0$\farcs$47 $\times$ 0$\farcs$45 (PA = +86$\degr$), for SO(6$_6$--5$_5$). First contours of both the SO maps start from 5$\sigma$ (35 mJy km s$^{-1}$ beam$^{-1}$, Upper, 25 mJy km s$^{-1} $ beam$^{-1}$, Lower) with intervals of 10$\sigma$. The orange thick contour is the CS(5--4) emission (25$\sigma$) which traces the outflow cavity walls associated with VLA1623A1+A2 \citep[from][]{Ohashi2022}. In magenta we plot selected contours from the high spatial resolution ($\sim$ 0$\farcs$2) continuum (0.9 mm) ALMA map by \citet{Harris2018} to pinpoint the positions of A1, A2, and B.} \label{fig:S} 
\end{center}
\end{figure}

The analysis of kinematics allows us to disclose
different molecular components emitting at different
velocities. Figure \ref{fig:S} shows the VLA1623-2417 A1+A2+B system as traced by both SO(5$_6$--4$_5$), and SO(6$_6$--5$_5$) emission integrated over $\pm$ 1 km s$^{-1}$ (velocity range labelled S, see Fig. \ref{fig:spectraSO}) with respect to the systemic velocity of the triple system of +3.8 km s$^{-1}$ \citep{Ohashi2022}. The emission at systemic velocity is mainly associated with the cavities, with additional features plausibly related with the VLA1623-2417 envelope. Figure \ref{fig:LV} shows the SO(5$_6$--4$_5$), and SO(6$_6$--5$_5$) maps of the blue-shifted
(by 1--6.6 km s$^{-1}$ with respect to $V_{\rm sys}$)
and red-shifted emission (by 1--2.5 km s$^{-1}$), i.e.
the intervals labelled as LV
in Fig. \ref{fig:spectraSO}. Note that the blue- and red-shifted LV ranges are asymmetric with respect to $V_{\rm sys}$ because they have been defined a posteriori after inspecting the SO dataset to identify velocity ranges tracing the same molecular structure. On these maps the 
intensity-weighted velocity CS(5--4) map (moment 1 map), by \citet{Ohashi2022}, is overlapped. The CS map reveals the rotation of the outflow cavity, with the southern sides red-shifted. 

The red-shifted SO LV emission is quite compact, as
highlighted in the zoom-in of Figure \ref{fig:LV}. The emission peaks towards the B protostar, plus an additional component starting
at the position of A1+A2 and inclined towards the SE direction, in agreement with the red-shifted outflow cavity \citep{Ohashi2022}.

On the other hand, the blue-shifted SO LV emission is very extended and clearly reveals a long ($\sim$ 1500 au) southern streamer %flowing from the outer regions of VLA1623-2417, and 
pointing to the central protostellar A+B system. 
Note that (i) the association with the outflow cavity is excluded from both the curved morphology, and, most importantly, (ii) by the fact that the outflow cavity in the southern region is red-shifted. 
These findings are well summarised by 
Fig. \ref{fig:PV}, which shows
the Position-Velocity (PV) cut (obtained with a slice width equal to the beam) of SO(5$_6$--4$_5$), black, and CS(5--4), magenta \citep{Ohashi2022},
along the southern direction (PA $ =0\degr$) from the position of VLA1623 A (upper panel) and VLA1623 B (lower panel).
The emission from the molecular cavity and the streamer are
located in different positions of the PV plot.

Crucial information on the streamer kinematics is also provided by Fig. \ref{fig:LVblue}, which shows, for the blue-shifted LV emission of both SO lines: the moment 1 image as well as the intensity-weighted velocity dispersion map (moment 2).
More precisely, the zoom-in region in the Right panels of Figure \ref{fig:LVblue} suggests that the streamer, 
once at $\sim$ 100 au from the protostars, is directing its gas mainly towards the B protostar, through an elongated feature well observed in the velocity dispersion map.
The moment 2 map also indicates that the velocity dispersion is higher towards B, in agreement with an inclination close to the edge-on geometry \citep[74$\degr$, ][]{Harris2018,Ohashi2022}.
Both the PV diagrams and the moment 1 maps show that the southern streamer is a coherent structure and slightly accelerating from $V_{\rm LSR}\sim 2$ km s$^{-1}$ at $-8\arcsec$ distance from the protostar VLA1623 B to $\sim 1.5$ km s$^{-1}$ at $-2\arcsec$ offset. 
This suggests that the streamer is conveying material towards the protostars.  

To summarise, the analysis of the spatial distribution and velocity of the SO emission indicate a streamer of gas extending from the outer envelope (out to 1500 au distance from A+B) to the central cluster plausibly feeding source B. The velocity and velocity dispersion increase towards the protostellar multiple system, possibly indicating accretion from the large scale envelope to the protostellar disks. Note that the streamer is blue-shifted, but it is on the side with red-shifted rotation of outflow and envelope \citep{Ohashi2022}. To make this happen, the streamer needs to infall from the backside of sources, and it will go behind the central sources.

\subsubsection{Northern region: the VLA1623 A accretion streamer}

\begin{figure*}
\begin{center}
\includegraphics[scale=0.65]{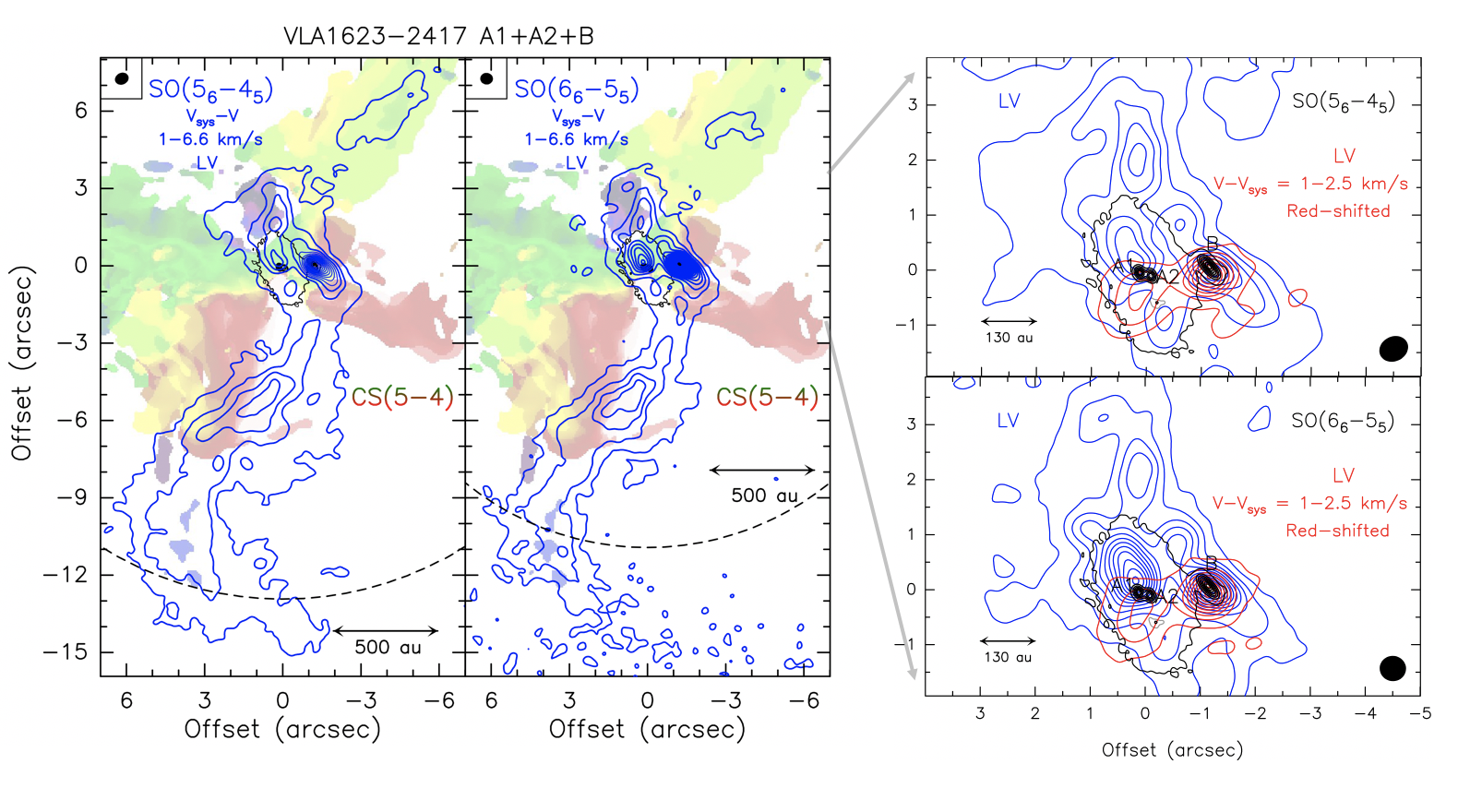}
\caption{The VLA1623-2417 A1+A2+B system as traced by SO(5$_6$--4$_5$) (Left panel), and SO(6$_6$--5$_5$) (Middle) emission blue-shifted by 1--6.6 km s$^{-1}$ and red-shifted by 1--2.5 km s$^{-1}$ (labelled LV, see Fig. \ref{fig:spectraSO}) with respect to the systemic velocity of the triple system of +3.8 km s$^{-1}$ \citep{Ohashi2022}. For sake of clarity the contours of the red-shifted spatial distribution are reported only in the zoom-in in the Right panels. The position of the A1+A2, and B protostars are labelled. The dashed circles delimitate the FWHM Field of View of the ALMA image: 26$\arcsec$ for SO(5$_6$--4$_5$), and 22$\arcsec$ for SO(6$_6$--5$_5$). The synthesised beam (top-left corners) are 0$\farcs$54 $\times$ 0$\farcs$45 (PA = --74$\degr$), for SO(5$_6$--4$_5$), and 0$\farcs$47 $\times$ 0$\farcs$45 (PA = +86$\degr$), for SO(6$_6$--5$_5$). 
 First contours of both the SO maps start from 5$\sigma$ (25 mJy km s$^{-1}$ beam$^{-1}$, blue, 15 mJy km s$^{-1}$ beam$^{-1}$, red) with intervals of 10$\sigma$. 
 Colour image represents the moment 1 spatial distribution of the molecular cavity as traced by CS(5--4) by \citet{Ohashi2022}:
 the cavities are rotating with the red-shifted
 emission coming from the southern arms, while the blue-shifted emission (here in green to avoid confusions with the SO blue contours) associated with the northern arms.
 In black we plot selected contours from the high spatial resolution ($\sim$ 0$\farcs$2) continuum (0.9 mm) ALMA map by \citet{Harris2018} to pinpoint the positions of A1, A2, and B.} \label{fig:LV} 
\end{center}
\end{figure*}

\begin{figure}
\begin{center}
\includegraphics[scale=0.55]{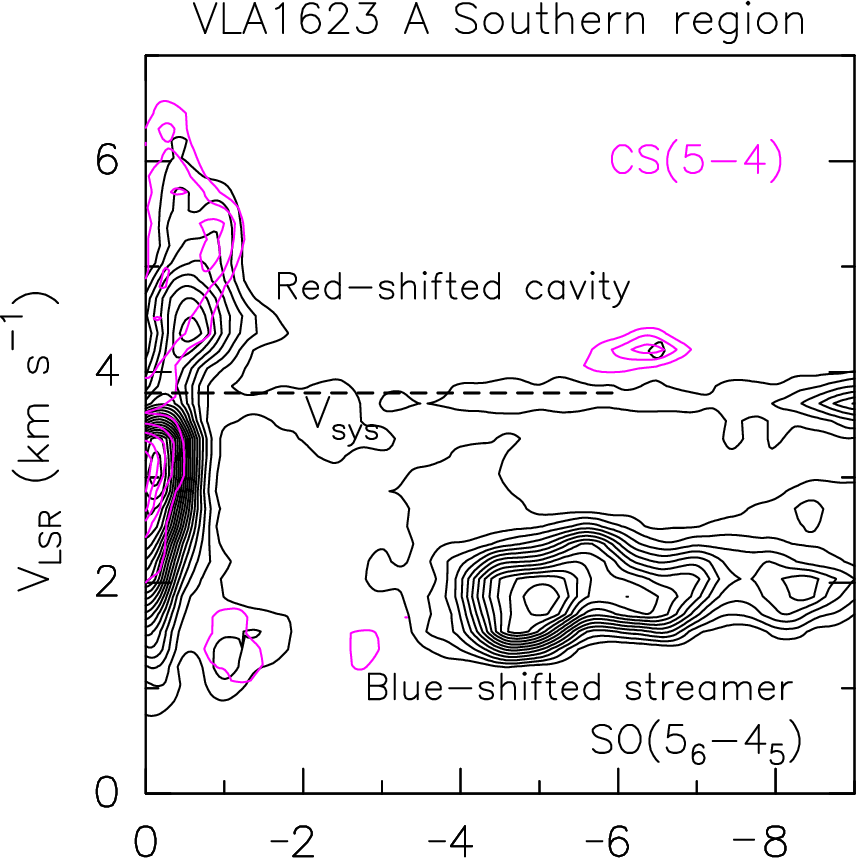}
\includegraphics[scale=0.55]{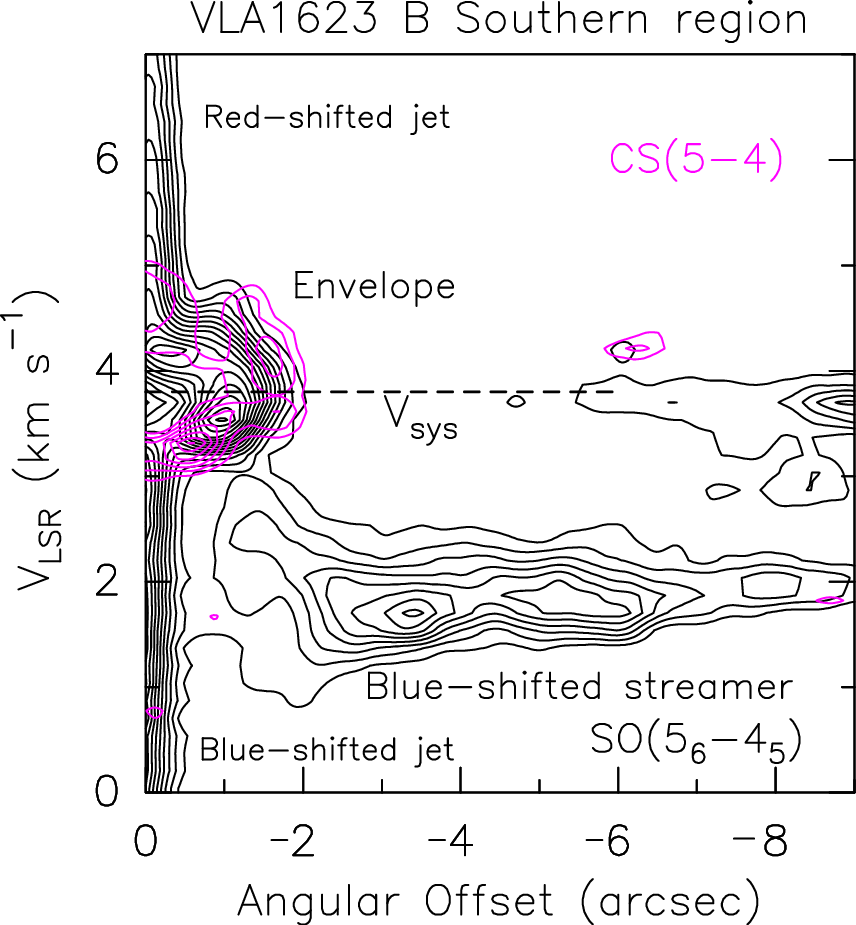}
\caption{Position-Velocity cut (beam averaged) of SO(5$_6$--4$_5$), black, and CS(5--4), magenta \citep{Ohashi2022},
along the southern direction (PA = 0$\arcsec$) centered on the position of VLA1623 A (Upper panel) and VLA1623 B
(Lower panel)}. Contour levels range from 5$\sigma$ (10 mJy beam$^{-1}$) by steps of 8$\sigma$. Dashed lines marks the systemic velocity 
\citep[+3.8 km s$^{-1}$,][]{Ohashi2022}.\label{fig:PV} 
\end{center}
\end{figure}

\begin{figure*}
\begin{center}
\includegraphics[scale=0.5]{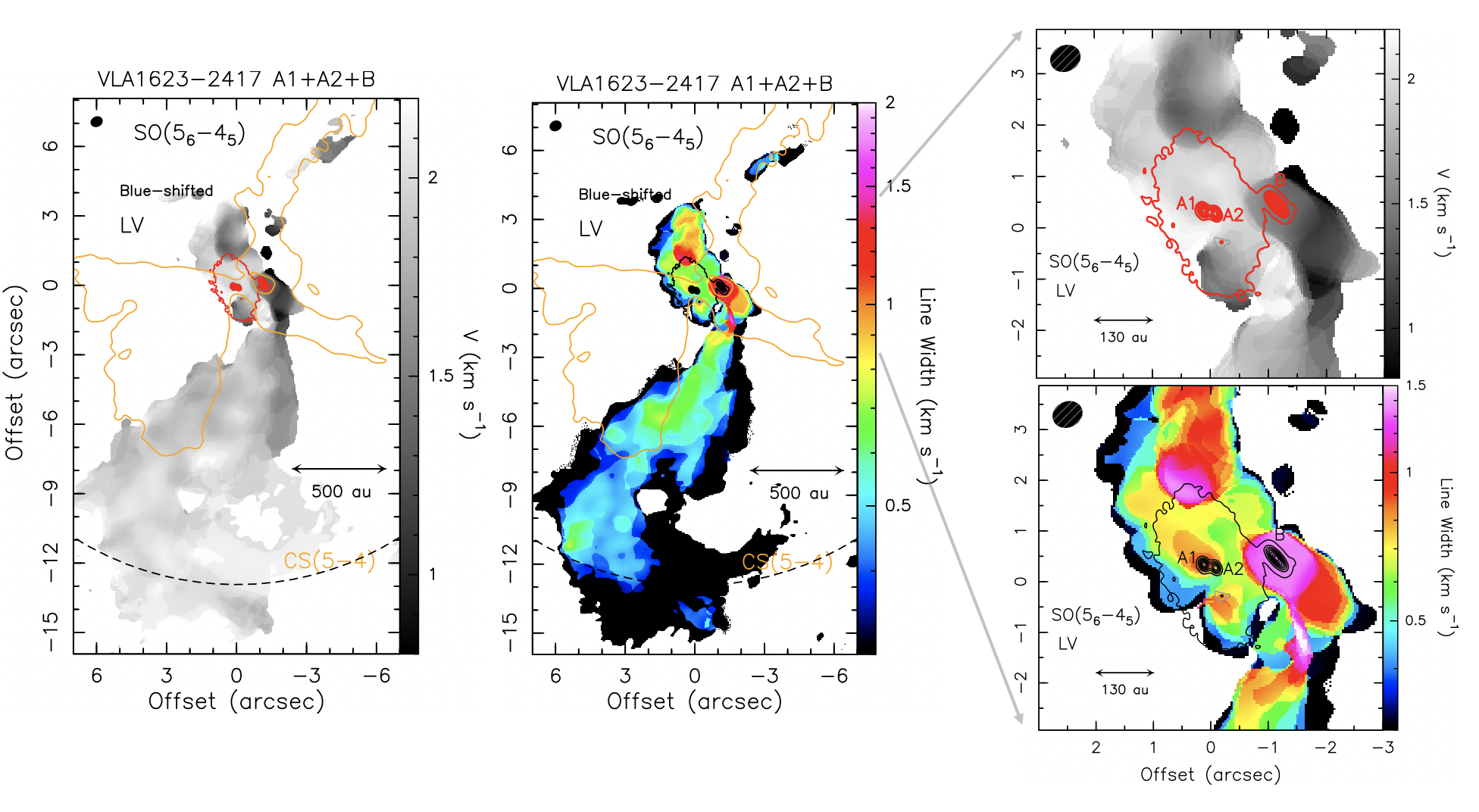}
\includegraphics[scale=0.5]{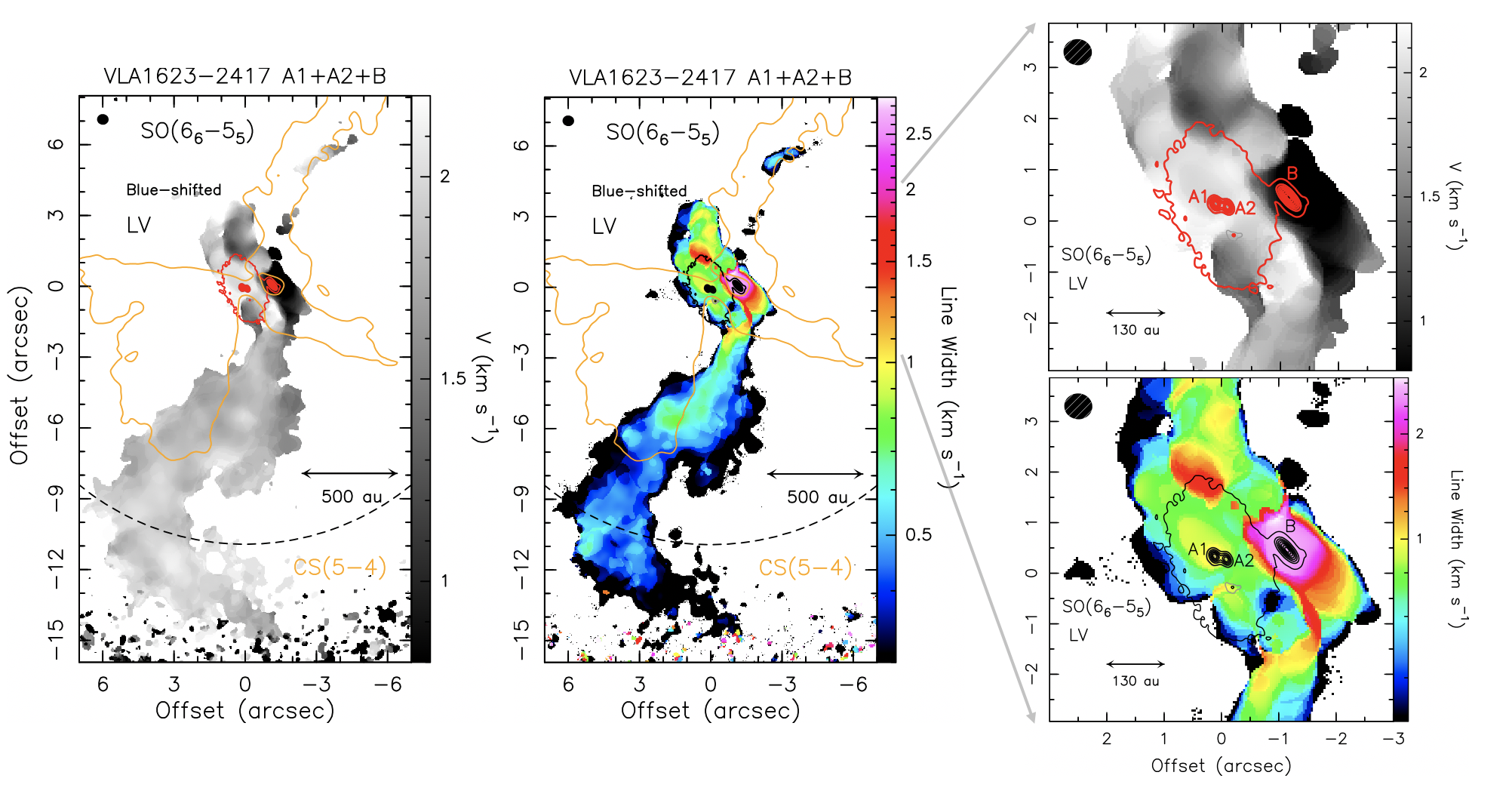}
\caption{Kinematics of the VLA1623--2417 A1+A2+B system as traced by the SO(5$_6$--4$_5$) (Upper panels), and SO(6$_6$--5$_5$) (Bottom panels) emissions {\it blue-shifted} with respect to systemic velocity
\citep[+3.8 km s$^{-1}$,][]{Ohashi2022} of 1--6.6 km s$^{-1}$ (labelled LV, see Fig. \ref{fig:spectraSO}). {\it Left and Middle panels} 
are for the moment 1 (intensity-weighted peak velocity), and moment 2 (intensity-weighted emission width) maps, respectively (colour scale).
First contours of both the SO maps start from 5$\sigma$ (25 mJy km s$^{-1}$ beam$^{-1}$).
The position of the A1+A2, and B protostars are labelled. The dashed circles delimitate the FWHM Field of View of the ALMA image: 26$\arcsec$ (Upper), and 22$\arcsec$ (Bottom). The synthesised beam (top-left corners) are 0$\farcs$54 $\times$ 0$\farcs$45 (PA = --74$\degr$), for SO(5$_6$--4$_5$), and 0$\farcs$47 $\times$ 0$\farcs$45 (PA = +86$\degr$), for SO(6$_6$--5$_5$). In red or black we plot selected contours from the high spatial resolution ($\sim$ 0$\farcs$2) continuum (0.9 mm) ALMA map by \citet{Harris2018} to pinpoint the positions of A1, A2, and B. The orange thick contour is the CS(5--4) emission (25$\sigma$) which traces the outflow cavity walls associated with VLA1623A1+A2
\citep[from][]{Ohashi2022}. {\it Right panels}: Zoom-in of the inner region around the circumbinary A1+A2 disk and the protostellar B disk.} \label{fig:LVblue} 
\end{center}
\end{figure*}

Focusing on the region north of A+B, 
Figure \ref{fig:LV} shows two small ($\sim$ 1$\arcsec$) elongated features, which could be associated with the blue-shifted cavity expected in these regions, plus a longer ($\sim$ 2$\arcsec$) structure located along the
N-S direction (see the zoom-in in the right panels). 
The latter is not spatially associated with the outflow cavity, therefore plausibly being an accretion streamer, in agreement with what \citet{Hsieh2020} proposed using the SO(8$_8$--7$_7$) line. Again, instructive information is provided by kinematics. Figure \ref{fig:LVblue} 
shows that the northern LV streamer has an increase of the intensity-weighted emission line width 
coinciding (on the plane of the sky) with the outer regions of the circumbinary disk around A1+A2. In conclusion, these findings are very suggestive 
that material falls onto the circumbinary disk at the position where SO emission is broader. No further information on the fate of the
material of the circumbinary disk is learned from the present data.

\subsection{SO and SiO jet emission}

\begin{figure*}
\begin{center}
\includegraphics[scale=0.5]{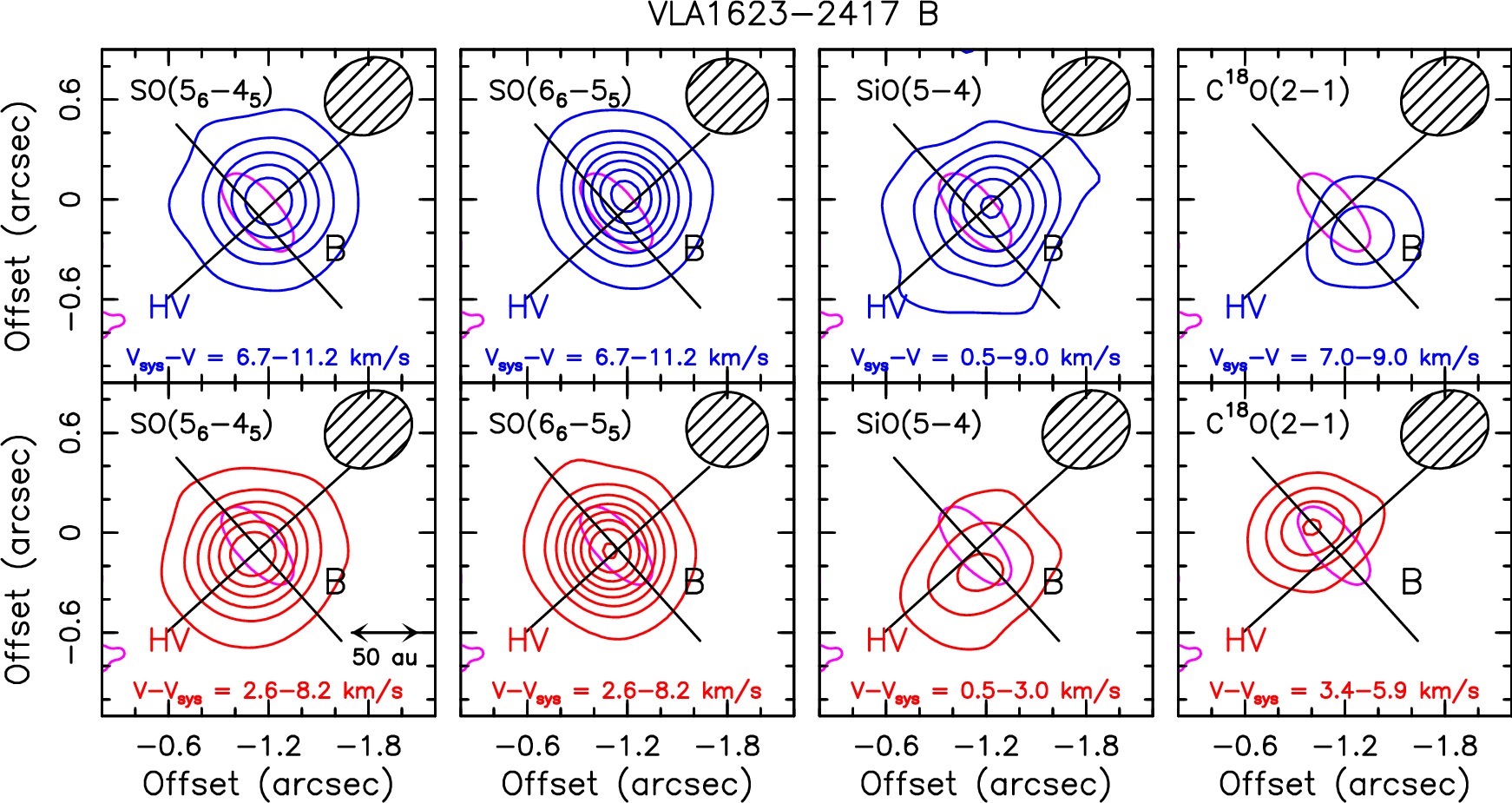}
\caption{Kinematics of the VLA1623--2417 B protostar as traced by the SO(5$_6$--4$_5$), and
SO(6$_6$--5$_5$) {\it (Left panels)} emission at the highest velocities with respect to systemic velocity
\citep[+3.8 km s$^{-1}$,][]{Ohashi2022}: blue-shifted by up to 11.2 km s$^{-1}$ {\it (Upper panels)}, and red-shifted by up to 8.2 km s$^{-1}$ {\it (Lower panels)}.
These velocity ranges are labelled as HV in Fig. \ref{fig:spectraSO}.
First contours of both the SO maps start from 5$\sigma$ (20 mJy km s$^{-1}$ beam$^{-1}$) by steps of 10$\sigma$. The synthesised beam (top-right corners) are 0$\farcs$54 $\times$ 0$\farcs$45 (PA = --74$\degr$), for SO(5$_6$--4$_5$), and 0$\farcs$47 $\times$ 0$\farcs$45 (PA = +86$\degr$), for SO(6$_6$--5$_5$). In magenta we plot a selected contour from the high spatial resolution ($\sim$ 0$\farcs$2) continuum (0.9 mm) ALMA map by \citet{Harris2018}. The tilted black cross indicates the disk inclination (PA = 42$\degr$) and the normal direction expected for the jet axis. 
{\it (Right panels)}: SiO(5--4) and C$^{18}$O(2--1) blue- and red-shifted emission. The
SiO(5--4) line is weaker than the SO ones: first contours and steps correspond to 3$\sigma$: 9 mJy km s$^{-1}$ beam$^{-1}$),
and 6 mJy km s$^{-1}$ beam$^{-1}$ for the blue- and red-shifted emission, respectively. The velocity ranges are smaller fo SiO (see text), while for C$^{18}$O the highest velocities tracing emission around B have been selected and reported in the labels. The beam is that of the SO(5$_6$--4$_5$) image.} \label{fig:HV}  
\end{center}
\end{figure*}

Figure \ref{fig:HV} shows the SO spatial distribution at the highest velocities with respect to $V_{\rm sys}$ = 
+3.8 km s$^{-1}$: blue-shifted by up to 11.2 km s$^{-1}$, and red-shifted by up to 8.2 km s$^{-1}$. 
This velocity range has been labelled as HV in 
Fig. \ref{fig:spectraSO}. 
Note that the disk size derived from the high-spatial resolution continuum  by \citet{Harris2018} is plotted in magenta. Both the SO(5$_6$--4$_5$)
and SO(6$_6$--5$_5$) emissions are compact and overlap with the position of the protostar B. 
The red-shifted and blue-shifted emission peaks are spatially separated, and  located along the SE-NW direction. This direction is perpendicular to the disk position angle  \citep[42$\degr$, ][]{Harris2018}.
In turn, these findings support the association of HV SO
with outflowing motion driven by VLA1623 B.  The velocities once deprojected using the geometry of the protostellar system \citep[disk inclination $\simeq$ 74$\degr$,][]{Harris2018}
reaches values $\sim$ 40 km s$^{-1}$ with respect to the systemic velocity.

The SiO(5--4) line has been detected, for the first time, in the VLA1623 star forming region. 
Fig. \ref{fig:mom0} shows the moment 0 map: the emission is in fact spatially unresolved and it  overlaps on the position of the B protostar. The spectrum  towards VLA1623 B is shown in Fig. \ref{fig:spectraSOSiO}: the line
is centred at the systemic velocity (+3.8 km s$^{-1}$), and it extends up to about +6 km s$^{-1}$ and down to +2 km s$^{-1}$. 
Figure \ref{fig:HV} shows the blue- and red-shifted SiO emission: as for SO at the highest velocities, SiO is associated with a velocity gradient, with
the red-shifted emission spatially offset towards SE (with respect to the continuum emission), while the blue-shifted emission peaks at NW.
As a typical high-velocity shock tracer, SiO then probes the protostellar jet driven by VLA1623 B. This is consistent with the CO(2--1) ALMA Band 6 images by \citet{Santangelo2015}: their maps have a lower spatial resolution (0$\farcs$65) than the FAUST dataset, but they indicate the same spatial offset for emission at velocities blue- and red- shifted by at least 6 km s$^{-1}$. 
The SiO radial velocities are lower than for SO. This could be due to the fact that the SO emission probes a wider angle layer of the wind with respect to SiO, which is expected to probe the inner collimated jet portion, as seen, e.g., in the high resolution ALMA maps of HH 212 \citep[see e.g.][]{Lee2019}. In this scenario the SiO gas would lie closer to the plane of the sky, which would explain lower observed radial velocities. 
Moreover, the estimated jet velocity could be a lower limit since it is obtained by deprojecting the SiO and SO radial velocity for the inclination derived for the disk ($\sim 74\degr$). The estimate of disk inclination for systems that are close to edge-on is affected by large uncertainty \citep[e.g.][]{Villenave2020}, and an inclination of larger than 85$\degr$ would lead to a typical jet velocity of at least $100$ km s$^{-1}$ \citep{Podio2021}.
Finally, note that the direction of the SiO velocity gradient is perpendicular (within the present spatial resolution) to the rotating
protostellar disk recently traced using methanol by \citet{Codella2022} and at the C$^{18}$O(2--1) HV emission
(here traced in Fig. \ref{fig:HV}). 
This comparison again supports that SiO traces the protostellar jet ejected by VLA1623 B.

%\subsection{The W protostar as imaged by SO}

%A special case is represented by the VLA1623 W object, which lies at 
%$\sim$ 1400 au from the A+B system (see Fig. \ref{fig:continuum}), associated with a systemic velocity
%of +1.6 km s$^{-1}$ (Mercimek et al. 2023).
%It is still debated whether VLA1623 W has been ejected from the VLA1623 A+B triple system or if it gravitationally bound to the VLA1623 cluster.
%As drawn in Figure \ref{fig:spectraSOSiO}, VLA1623 W is associated with SO emission, while no SiO emission is detected. The SO line profiles
%are asymmetric with a peak blue-shifted by $\sim$ 2 km s$^{-1}$) with respect to the systemic velocity.
%Figure \ref{fig:W} shows the kinematics of VLA1623--2417 W as traced by 
%both the SO(5$_6$--4$_5$) and SO(6$_6$--5$_5$) lines.
%The blue- and red-shifted emissions peaks of the blue- and red-shifted emission are located along the disk major axis (see the black contours for dust continuum) and they are spatially offset, with a velocity gradient along the N-S direction.
%Interestingly, this effect is enhanced at velocities blue- and red-shited (with respect to +1.6 km s$^{-1}$) by at least 2 km s$^{ -1}$: at the highest blue and red velocities the emisson peaks are better spatially split. In other words, the emission is more compact and peaks at
%smaller distances for increasing velocities, as expected in a Keplerian rotating disk or for the inner portion of the rotating envelope. These findings are well in agreement with the recent observations
%of C$^{18}$O(2--1) by Mercimek et al. (2023), obtained in the FAUST context. 

\section{Discussion} \label{sec:discussion}

\begin{figure*}
\begin{center}
\includegraphics[scale=0.6]{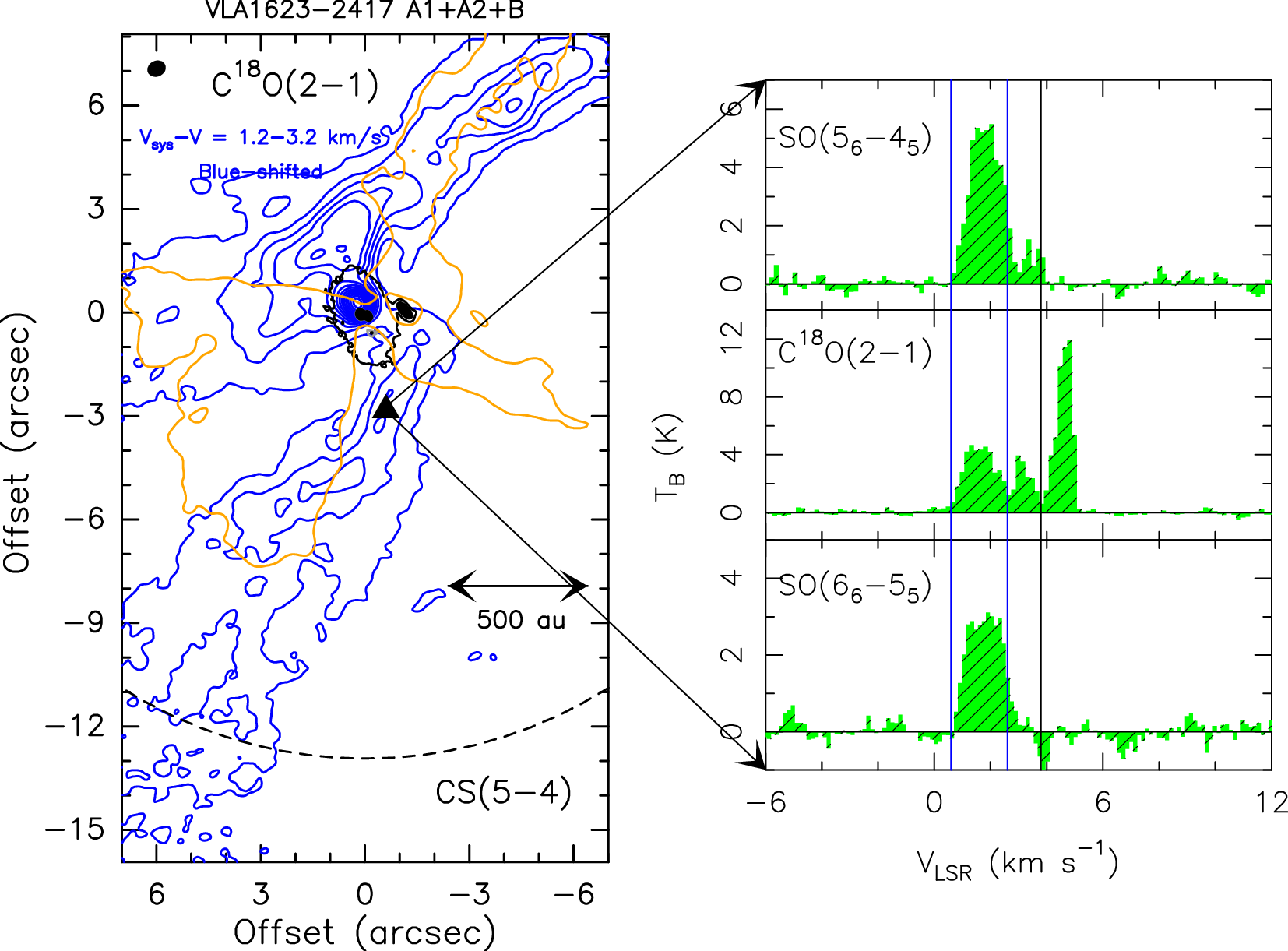}
\caption{{\it Left panel:} The VLA1623--2417 southern molecular streamer as traced by the
C$^{18}$O(2--1) emission at velocities blue-shifted by 1.4--3.2 km s$^{-1}$ with respect to $V_{\rm sys}$ = +3.8 km s$^{-1}$. The velocities are those tracing, in C$^{18}$O, the blue-shifted streamer (see text). 
First contours start from 5$\sigma$ (30 mJy km s$^{-1}$ beam$^{-1}$) with intervals of 3$\sigma$. 
{\it Right panels} SO(5$_6$--4$_5$), C$^{18}$O(2--1), and SO(6$_6$--5$_5$) spectra (in brightness temperature, $T_{\rm B}$, scale) derived at the position (--0$\farcs$7,--3$\farcs$0) associated with the streamer, and marked with a triangle in the Left panel. The black vertical lines are for the systemic velocity. The blue vertical lines delimit the velocities of the C$^{18}$O used to obtain the image of the southern streamer shown in the Left panel.} \label{fig:c18o} 
\end{center}
\end{figure*}

\subsection{Excitation temperature of the VLA1623 B accretion streamer}

In light of the SO results, and in order to constrain the physical parameters of the molecular streamer detected in the VLA1623 A+B region, we inspected the C$^{18}$O(2--1) dataset, published by \citet{Mercimek2023}.
Figure \ref{fig:c18o} (Left panel) shows the C$^{18}$O(2--1) map integrated over the velocities tracing the blue-shifted streamers, namely shifted by 1.4--3.2 km s$^{-1}$ with respect to $V_{\rm sys}$. 
The southern streamer accreting towards VLA1623 B is well
revealed, while in the northern portion of the map there is a clear contamination with the blue-shifted outflow cavity. We then proceeded to analyse the southern VLA1623 B streamer.
We extracted the spectrum at the position offset by --0$\farcs$7,--3$\farcs$0 (with respect to the phase center of the map, see Sect. 2), i.e. at the emission peak closest to the A+B system. The C$^{18}$O(2--1) line profile shows the Gaussian-like component associated with the accretion streamer.
%, as well as other lines at higher velocities tracing additional envelope components whose analysis is out of the scope of this paper. 
Figure \ref{fig:c18o} (Right panels) compares
the C$^{18}$O(2--1) lines with those of both the SO lines, extracted at the same position of the map. Also the SO spectra
show a Gaussian-like profile similar to that of C$^{18}$O.
Assuming an LTE (Local Thermodynamic Equilibrium) population and optically thin lines, a crude estimate of the SO excitation temperature ($T_{\rm ex}$) can be derived from the two SO observed lines: 33$\pm$9 K. To our knowledge, this is the first $T_{\rm ex}$ estimate of a molecular streamer based on two lines of the same species, being usually detected with one emission line of a molecular species \citep[see the recent review by][]{Pineda2023}.
Based on a simple toy model where the gas and dust are heated by the central protostars (without considering the
outflow cavities and the disks) \citep[e.g.][]{Ceccarelli2000}, we estimate the expected gas temperature at  $\sim$ 390 au distance from the protostars  
(where the spectra have been extracted). For a total bolometric luminosity of $\sim$ 1 $L_{\rm \sun}$, we find that the temperature is $\sim$ 20 K.
The estimated excitation temperature is higher, being in the 
24--42 K range. However, the comparison has to be taken with a pinch of salt, being based on two transitions: more lines need to be observed to investigate the reliability of the LTE assumption, as well as possible line opacity effects. In addition, (i) 
if the emission is thermalised, the temperature is likely to increase near the cavity walls, being thus closer to the SO excitation temperature, and (ii) there are the uncertainties due to projection effects and to the length of the material along the line of sight.
%\citet{Valdivia2022} assumed for the C$^{18}$O streamer observed in the Class I Per-emb-50 an excitation temperature of 15$\pm$5 K, based on previous measuremnts of the molecular protostellar envelope.
%...35-45 K for the DCN and C$^{18}$O streamer of the Class I SVS13-A binary, derived using the protostellar luminosity and the distance of the streamer from the SVS13-A protostars \citep{Hsieh2023}. 

Note that the excitation temperature measured towards the SO region where the northern streamer impacts with the circumbinary VLA1623 A disk (see Fig. \ref{fig:LVblue} at +0$\farcs$3,+1$\farcs$5 from the map center) is higher than the value measured in the southern streamer, 55$\pm$12 K, a temperature plausibly increased due
to a slow shock at the impact location. 
The SO excitation temperature has been estimated also
at the position where the southern streamer seems to impact onto 
the disk of the B protostar (see Fig. \ref{fig:LVblue} at --1$\farcs$4,--0$\farcs$2): the temperature is high, 63$\pm$12 K, and it can be explained again by a shock.
Alternatively, given the proximity of the position to B, the high temperature could be due to protostellar heating.
Again, this has to be verified using multiple SO lines for a more reliable temperature measurement.

\subsection{Accretion and infalling rates}

At a temperature of 33 K, the total SO column density is $N_{\rm SO}$ $\simeq$ 2 $\times$ 10$^{14}$ cm$^{-2}$. To derive the uncertainty, $N_{\rm SO}$ increases by a factor 2 assuming 20 K instead of 33 K. The total column density of C$^{18}$O is 4 $\times$ 10$^{15}$ cm$^{-2}$. Using the classical $^{16}$O/$^{18}$O = 560 and CO/H$_2$ = 10$^{-4}$ \citep{Wilson1994}, 
the H$_2$ total column density is 2 $\times$ 10$^{22}$ cm$^{-2}$. The total mass of the blue-shifted southern streamer can be estimated from the emitting region and the estimate of the average C$^{18}$O (and consequently H$_2$) column density throughout the emitting region: M$_{\rm streamer}$ $\simeq$ 3 $\times$ 10$^{-3}$ $M_{\rm \sun}$. 
This estimate is lower with respect to the total mass of the HC$_3$N long (10$^{4}$ au) streamer detected by \citet{Pineda2020} towards the Class 0 
object IRAS 03292+3039 (M$_{\rm streamer}$ = 0.1--1 $M_{\rm \sun}$). On the other hand,
if we compare the VLA 1623--2417 southern streamer with the 
Class I streamers, our estimates are similar: 
SVS13-A \citep[4 $\times$ 10$^{-3}$ $M_{\rm \sun}$,][]{Hsieh2023} and Per-emb-50 \citep[1 $\times$ 10$^{-2}$ $M_{\rm \sun}$,][]{Valdivia2022}. 
%It would be tempting to correlate the mass streamer with the evolutionary stage of the acrreting protostars. However, beside the evident lack of statistics, the comparison between the total mass of the streamers strongly depends on its length, which
%in all these cases looks not fully revovered because larger than the FoV of the interferometric (IRAM-NOEMA, ALMA) images. 

As the southern streamer is impacting on the disk of source VLA1623 B, we aim to compare the mass infall rate of the streamer with the mass accretion rate on source B, to understand how much streamers may contribute to set the final mass of protostellar objects. This is indeed still an open question,
given the paucity of measurements of the physical properties of accretion streamers.
On the one hand, \citet{Pineda2020}
and \citet{Valdivia2022} found that the accretion rates of the streamers in IRAS 03292+3039 and Per-emb-50, are of the same order of magnitude
of the protostellar accretion rates. On the other hand, \citet{Hsieh2023}, found an accretion rate of the streamer lower by an order of magnitude with respect to the protostellar accretion in the SVS13-A source. An estimate of the free-fall timescale of the southern streamer accreting VLA1623 B
can be obtained using the classical equation \citep[e.g][]{Pineda2020,Pineda2023},

\begin{equation}
t_{\rm ff} = \sqrt{R^3/GM_{\rm total}},
\end{equation}

\noindent
where R is the streamer length,
$M_{\rm total}$ is the mass inside R, and G is the gravitational constant. Taking R = 1500 au, 
a total mass in the 1--2 $M_{\rm \sun}$ range
\citep[e.g.][]{Murillo2018L,Ohashi2022}, we obtain, for the southern blue-shifted streamer, $t_{\rm ff}$ $\simeq$ 6--9 $\times$ 10$^{3}$ yr. 
Note that the free-fall velocity lies in the range 0.9--1.3 km s$^{-1}$,
i.e. values quite close (56\%--81\%) to the difference in velocity, 1.6 km s$^{-1}$, observed within the southern streamer.
By dividing the streamer mass with the free-fall timescale we obtain an estimate of the accretion rate of the southern streamer onto the B protostar: 3--5 $\times$ 10$^{-7}$ $M_{\rm \sun}$ yr$^{-1}$. 

To estimate the mass accretion rate on source B, we assume that the source bolometric luminosity is due to the gravitational energy released by the accretion onto the protostar ($L_{\rm bol}$ = $L_{\rm acc}$), 
and estimate the mass accretion as: 
$\dot{M}_{\rm acc}$ = $L_{\rm bol}$$R_{\rm *}$/G$M_{\rm *}$. 
The bolometric luminosity of source B derived by 
\citet{Murillo2018L} based on the source spectral energy distribution is 0.2--0.3 $L_{\rm \sun}$, while the protostellar mass has been estimated from the fit of the rotation curve of the disk by \citet{Ohashi2022}, giving a dynamical mass of 1.7 $M_{\rm \sun}$. 
Based on these values, and assuming that the stellar radius is $R_{\rm *}$ = 2 $R_{\rm \sun}$ \citep{Stahler1988} we infer $\dot{M}_{\rm acc}$ = 10$^{-8}$ $M_{\rm \sun}$ yr$^{-1}$.
The estimated mass accretion rate is highly uncertain because it depends strongly on the protostellar properties, which may be affected by large uncertainties, and because accretion may be episodic and characterized by accretion bursts \citep{Fisher2023}. 
In particular, the estimated dynamical mass is uncertain, due to the intermediate angular resolution of the FAUST data \citep[50 au,][]{Ohashi2022}.
If we assume the typical range of masses kinematically estimated for low-mass protostellar objects, i.e. $M_{\rm *}$ = 0.05--0.25 $M_{\rm \sun}$
\citep{Choi2010,Kwon2015,Yen2017,Lee2020}, we obtain a mass accretion rate up to 6 $\times$ 10$^{-8}$ 
$M_{\rm \sun}$ yr$^{-1}$ (for 0.25 $M_{\rm \sun}$) and 3 $\times$ 10$^{-7}$ $M_{\rm \sun}$ yr$^{-1}$ (for 0.05 $M_{\rm \sun}$).
In summary, as the streamer infall rate is about 
3--5 $\times$ 10$^{-7}$ $M_{\rm \sun}$ yr$^{-1}$ the mass fed by the streamer is comparable with the total mass accretion rate. 
%Regarding the funnelling of fresh material, \citet{Kufmeier2023} has been exploring how late infall can impact the disc angular momentum budget and size, ad also how it can {\it rejuvenate} protostars. In particular, two aspects should be noted: (i) a) the centrifugal radius is higher for stars with substantial infall, suggesting that large disks are the result of recent infall events, and (ii) some late accretors could be inaccurately categorised as Class 0 simply because they are embedded. 

\subsection{SO abundances in the southern VLA1623 B streamer}

The SO abundance relative to H$_2$ can be derived
for the LV southern streamer by comparing the SO and H$_2$ column densities extracted at the (--0$\farcs$7,--3$\farcs$0) position, where C$^{18}$O emission is dominated by the streamer emission (see Sect. 5.1):
$X_{\rm SO}$ $\simeq$ 10$^{-8}$. This value is 
larger than that measured in the gas phase in molecular clouds located in Perseus, Taurus, and Orion \citep[0.7--2 $\times$ 10$^{-9}$,][and references therein]{Navarro-Almaida2020,Rodriguez2021}.
On the other hand $X_{\rm SO}$ $\simeq$ 10$^{-8}$ is
at the lower end of the SO abundance range derived for hot-corinos around protostars up to $\sim$ 10$^{-7}$ \citep[e.g.][and references therein]{Codella2021}.
However, the hot-corino nature, i.e. the thermal 
evaporation of the dust mantle in the streamer, is here excluded (assuming LTE conditions), considering the derived excitation temperature of $\sim$ 30 K.
Even the occurrence of strong shocks ($V_{\rm shocks}$ $\geq$ 10 km s$^{-1}$) has to be excluded 
given that they would increase the SO abundance up
to higher values than those observed in the southern streamer \citep[$\sim$ 10$^{-7}$, e.g.][]{Bachiller1997,Bachiller2001,Feng2020}. 
A possibility to explain an SO abundance larger than 
those typical in starless molecular clouds is to speculate the occurrence of mild shocks ($V_{\rm shocks}$ of a few km s$^{-1}$), induced by
the accretion of the gas through the streamer, releasing
part of the Sulphur on dust mantles. 
Interestingly, \citet{vanGelder2021} modeled
the Sulphur chemistry in low-velocity shocks (down
to $\sim$ 3--4 km s$^{-1}$), showing that SO can be efficiently formed from SH reacting with the O atom and/or S with OH.
The SO chemistry in the streamer could mimic that
observed in the L1448-mm protostar \citep{Jimenez2005},
where the weak shock precursor component 
increases the SO abundance by one order of magnitude only.

%\begin{figure*}
%\begin{center}
%\includegraphics[scale=0.45]{codella-2019258-LV2mom0mom1mom2.png}
%\caption{Kinematics of the VLA1623--2417 A1+A2+B system as traced by the SO(6$_6$--5$_5$) and SO(5$_6$--4$_5$) emission {\it blue-shifted} with respect to systemic velocity
%\citep[+3.8 km s$^{-1}$,][]{Ohashi2022} of 2.6--6.6 km s$^{-1}$ (velocity range labelled HV, see Fig. \ref{fig:spectraSO}.
%{\it Left panels} are for the moment 0 image (colour scale).
%First contours of both the SO maps start from 5$\sigma$ (23 mJy km s$^{-1}$ beam$^{-1}$), with steps of 10$\sigma$.
%{\it Middle and Right panels} are for the momentum 1
%(intensity-weighted peak velocity), and moment 2
%(intensity-weighted emission width) maps, respectively (colour scale).
%The position of the A1+A2, B, and W protostars are labelled. The synthesised beam (top-left corners) is 0$\farcs$47 $\times$ 0$\farcs$45 (PA = +86$\degr$), for SO(6$_6$--5$_5$). First contours of both the SO maps start from 5$\sigma$ (20 mJy km s$^{-1}$ beam$^{-1}$). In black we plot selected contours from the high spatial resolution ($\sim$ 0$\farcs$2) continuum (0.9 mm) ALMA map by \citet{Harris2018} to pinpoint the positions of A1, A2, and B. The orange thick contour is the CS(5--4) emission (25$\sigma$) which traces the outflow cavity walls associated with VLA1623A1+A2
%\citep[from][]{Ohashi2022}.} \label{fig:LV2blue} 
%\end{center}
%\end{figure*}

\subsection{VLA 1623B: the SiO jet}

Here we estimate the beam-averaged column density in the HV SO component (see Fig. \ref{fig:spectraSO}) as well as of the SiO jet. 
Assuming LTE conditions and optically thin emission, the excitation temperature of the HV SO ranges between 92$\pm$18 K
(emission red-shifted by up to 8.2 km s$^{-1}$) and 102$\pm$19 K (emission blue-shifted by up to 11.2 km s$^{-1}$).
This supports the association of HV SO with shocked regions created by the propagation of the jet driven by VLA1623 B, as observed in several
protostellar regions \citep[e.g.][and references therein]{Bachiller2001,Taquet2020,Feng2020,Podio2021}. With these temperatures the SO column density is $\sim$ 5 $\times$ 10$^{14}$ cm$^{-2}$.
The SiO total column density has been derived assuming
a typical jet temperature of 100$\pm$50 K \citep[e.g.][]{Podio2021}, obtaining $N_{\rm SiO}$ = 2--5 $\times$ 10$^{12}$ cm$^{-2}$. Unfortunately, the SO and SiO abundances cannot be constrained because the C$^{18}$O(2--1)
emission at these highest detected velocities (up to 6 km s$^{-1}$ and down to 9 km s$^{-1}$ with respect to $V_{\rm sys}$) are
tracing a compact structure rotating along a direction perpendicular to the SiO jet axis (Fig. {\ref{fig:HV}}, Right panels). As a matter of fact, C$^{18}$O observed on spatial scales below 100 au is an efficient tracer of the inner protostellar envelope and/or accretion disks \citep{Murillo2015,Bergner2019a,Zhang2021,Mercimek2023}. In the VLA1623 B case, C$^{18}$O(2--1) traces the same rotating gas observed as the CH$_3$OH
by \citet{Codella2022} using the FAUST dataset.

%\subsection{VLA 1623W: the SO disk}

\section{Conclusions} \label{sec:conclusions}

In the context of the FAUST ALMA Large Program, the VLA1623-2417 protostellar 
cluster has been imaged at 1.2--1.3 mm in 
the SO(5$_6$--4$_5$), SO(6$_6$--5$_5$), and SiO(5--4) emissions at the spatial scale of 50 au. In particular, we focused on VLA1623 A and its circumbinary disk, and
on the VLA1623 B protostar.
The main findings are summarized as follows:

\begin{itemize}
    \item SO shows extended ($\sim$ 20$\arcsec$, 260 au) emission, peaking towards the A and B protostars, where the observed spectra are consistent with the association with the A and B hot-corinos. An absorption dip is present in the VLA1623 B profile. The absorption is more prominent for SO(5$_6$--4$_5$), suggesting the presence of a cold
    SO component along the line of sight;
    \item The analysis of the SO kinematics allows us to reveal different structures emitting at different velocities. At the systemic velocity (+3.8 km s$^{-1}$) elongated SO structures are associated with the outflow cavities previously imaged in CS. Velocities blue-shifted by 1--6.6 km s$^{-1}$ reveal a long ($\sim$ 2000 au) southern streamer, with an increase in the mean velocity of $\sim$ 1.6 km s$^{-1}$ approaching the central A+B system, and apparently feeding the VLA1623 B protostar. In addition, a $\sim$ 2$\arcsec$ (260 au) streamer, previously observed by \citet{Hsieh2020}, is imaged through the N-S orientation, impacting from North the A circumbinary disk;
    \item The SiO emission, detected for the first time in VLA1623-2417, is very compact ($\sim$ 100 au) and associated only with the B protostar. The HV SO emission, red- and blue-shifted up to $\sim$ 10 km s$^{-1}$ is also compact ($\leq$ 100 au), and overlaps with the B protostar, as shown by SiO(5--4). Assuming LTE conditions and 
    optically thin lines, an estimate of the HV SO excitation temperature can be derived: 92$\pm$5 K(red) and 102$\pm$6 K (blue), showing the association of HV SO with shocks created by VLA1623 B jet. Using these temperatures, the SO and SiO total column densities are $N_{\rm SO}$ = 5 $\times$ 10$^{14}$ cm$^{-2}$, and $N_{\rm SiO}$ = 2--5 $\times$ 10$^{12}$ cm$^{-2}$, respectively;
    \item An estimate of the SO excitation temperature of the southern streamer can also be derived
    (LTE, optically thin emission): 33$\pm$9 K. The total SO column density is 2 $\times$ 10$^{14}$ cm$^{-2}$. Using C$^{18}$O(2--1) FAUST data \citep{Mercimek2023}, we estimated the SO abundance: $X_{\rm SO}$ $\simeq$ 10$^{-8}$, a value
    higher than what is usually found in molecular clouds.
   We speculate the occurrence of weak shocks induced by the accretion through the shock which could release
   into the gas-phase part of the dust mantles;
\item 
The total mass of the blue-shifted southern streamer is 3 $\times$ 10$^{-3}$ $M_{\rm \sun}$. This estimate is in agreement with those observed in Class I objects: 4 $\times$ 10$^{-3}$ $M_{\rm \sun}$ for SVS13-A \citep{Hsieh2023}, and 1 $\times$ 10$^{-2}$ $M_{\rm \sun}$ for Per-emb-50 \citep{Valdivia2022}.
On the other hand, our estimate is 
lower with respect to that measured in
the Class 0 
object IRAS 03292+3039 \citep[0.1--1 $M_{\rm \sun}$,][]{Pineda2020}.
It would be tempting to correlate the streamer mass with the evolutionary stage of the acrreting protostars. However, beside the evident lack of statistics, the comparison between the total mass of the streamers strongly depends on its length, which looks not fully traced because it is larger than the FoV of the interferometric (IRAM-NOEMA, ALMA) images;
\item
The free-fall timescale of the southern streamer is 6--9 $\times$ 10$^{3}$ yr. Consequently, the estimate of the 
the accretion rate of the streamer on the B protostar is 3--5 $\times$ 10$^{-7}$ $M_{\rm \sun}$ yr$^{-1}$. 
This can be compared with the mass accretion rate, $\dot{M}_{\rm acc}$, on source B, calculated between 6 $\times$ 10$^{-8}$ $M_{\rm \sun}$ yr$^{-1}$ and 3 $\times$ 10$^{-7}$ $M_{\rm \sun}$ yr$^{-1}$.
In conclusion, the mass fed by the streamer is a significative fraction of the total mass accretion rate of VLA1623 B. 
\end{itemize}

\section{Epilogue: uniqueness of the VLA1623--2717 region} \label{sec:epilogue}

The ALMA high-sensitivity FAUST data contributed to chemically characterise the already well studied VLA1623--2417 star forming region, imaging: CS, CCH, and H$^{13}$CO$^{+}$ \citep{Ohashi2022}, CH$_3$OH, and HCOOCH$_3$ \citep{Codella2022}, C$^{18}$O \citep{Mercimek2023}, SO, and SiO (this paper). As matter of fact, CH$_3$OH, HCOOCH$_3$, and SiO have been detected for the first time in VLA1623--2417.
In addition, the FAUST papers enlighted the multiple processes at work
in shaping a multiple protostellar system A1+A2+B. More specifically, there are strong hints of misaligned accretion from the southern environment (this paper), and the possible hierarchical decay of the multiple stellar system, where the A1 and B protostellar disks are counter-rotating \citep{Ohashi2022,Codella2022}, molecular envelope and outflows show a misalignament rotation \citep{Ohashi2022}, and with the ejection of one member 
of such an unstable system in the NE direction \citep[VLA1623 W]{Mercimek2023}.

\section*{Acknowledgements}
We thank the anonymous referee for their comments and suggestions definitely improved the manuscript. This project has received funding from the EC H2020 research and innovation
programme for: (i) the project "Astro-Chemical Origins” (ACO, No 811312), (ii) the European Research Council (ERC) project “The Dawn of Organic
Chemistry” (DOC, No 741002), and (iii) the European Research Council (ERC) project Stellar-MADE (No. 101042275, project Stellar-MADE).  CC, LP, and GS acknowledge 
the PRIN-MUR 2020  BEYOND-2p (Astrochemistry beyond the second period elements, Prot. 2020AFB3FX),
the PRIN MUR 2022 FOSSILS (Chemical origins: linking the fossil composition of the Solar System with the chemistry of protoplanetary disks, Prot. 2022JC2Y93), the project ASI-Astrobiologia 2023 MIGLIORA
(Modeling Chemical Complexity, F83C23000800005), and the INAF-GO 2023 fundings
PROTO-SKA (Exploiting ALMA data to study planet forming disks: preparing the advent of SKA, C13C23000770005). 
GS also acknowledges support from the INAF-Minigrant 2023 TRIESTE ("TRacing the chemIcal hEritage of our originS: from proTostars to planEts”).
EB aknowledge the Deutsche Forschungsgemeinschaft (DFG, German Research Foundation) under Germany´s Excellence Strategy – EXC 2094 – 390783311.
DJ is supported by NRC Canada and by an NSERC Discovery Grant.
LL acknowledges the support of UNAM DGAPA PAPIIT grants IN112820 and IN108324, and CONAHCYT-CF grant 263356.
SBC was supported by the NASA Planetary Science Division Internal Scientist Funding Program through the Fundamental Laboratory Research work package (FLaRe).
IJ-S acknowledges funding by grants No. PID2019-105552RB-C41 and PID2022-136814NB-I00 from the Spanish Ministry of Science and Innovation/State Agency of Research MCIN/AEI/10.13039/501100011033 and by “ERDF A way of making Europe".
This paper makes use of the following ALMA data: ADS/JAO.ALMA\#2018.1.01205.L. ALMA is a partnership of ESO (representing its member states), NSF (USA) and NINS (Japan), together with NRC (Canada), MOST and ASIAA (Taiwan), and KASI (Republic of Korea), in cooperation with the Republic of Chile. The Joint ALMA Observatory is operated by ESO, AUI/NRAO and NAOJ. The National Radio Astronomy Observatory is a facility of the National Science Foundation operated under cooperative agreement by Associated Universities, Inc. \\

%\vspace{-0.5cm}

\section*{DATA AVAILABILITY} 
The raw data are available on the ALMA archive at the end of the proprietary period (ADS/JAO.ALMA\#2018.1.01205.L).

\vspace{-0.5cm}
\bibliographystyle{mnras}
\bibliography{main} % if your bibtex file is called example.bib
%%%%%%%%%%%%%%%%%%%%%%%%%%%%%%%%%%%%%%%%%%%%%%%%%%
%%%%%%%%%%%%%%%%% APPENDICES %%%%%%%%%%%%%%%%%%%%%

\appendix

\bsp	% typesetting comment
\label{lastpage}
\end{document}